
\newif\iffiginclude
\figincludefalse        
\input PHYZZX
\iffiginclude
\input psfig
\fi
\date={September1994}
\rightline {September, 1994}
\rightline {UPR-623-T}
\title {SUPERSYMMETRIC DYONIC BLACK HOLES IN KALUZA-KLEIN THEORY}
\author {Mirjam Cveti\v c\foot{E-mail address: cvetic@cvetic.hep.upenn.edu}
and Donam Youm\foot{E-mail address: youm@cvetic.hep.upenn.edu}}
\address {Physics Department \break
          University of Pennsylvania, Philadelphia PA 19104-6396}
\abstract
{We study supersymmetric, four-dimensional (4-d), Abelian charged
black holes (BH's) arising in $(4+n)$-d ($1 \le n \le 7$)
Kaluza-Klein (KK) theories. Such solutions, which
satisfy supersymmetric Killing spinor equations (formally satisfied
for any $n$) and saturate
the corresponding Bogomol'nyi bounds, can be obtained if and only if
the isometry group of the internal space is broken down to the
$U(1)_E \times U(1)_M$ gauge group; they correspond to dyonic BH's
with electric $Q$ and magnetic $P$ charges associated with {\it
different} $U(1)$ factors.
The internal metric of such configurations is diagonal with
$(n-2)$ internal radii constant, while the remaining two radii
(associated with the respective electric and magnetic
$U(1)$ gauge fields) and the 4-d part of the metric turn out to be
independent of $n$, \ie , solutions are effectively those
of supersymmetric 4-d BH's of 6-d KK theory.
For $Q\ne 0$ and $P\ne 0$, 4-d space-time has a null
singularity, finite temperature ($T_H\propto 1/\sqrt{|QP|}$)
and zero entropy. Special cases with either $Q=0$
or $P=0$ correspond to the supersymmetric 4-d BH's of 5-d KK
theory, first derived by Gibbons and Perry, which have a naked
singularity and infinite temperature. }

\endpage

\REF\BOG{E.B. Bogomol'nyi, Sov. J. Nucl. Phys. {\bf 24} (1976) 449.}

\REF\ADM{R. Arnowitt, S. Deser and C.W. Misner, Phys. Rev. {\bf 122}
(1961) 997; L.F. Abbott and S. Deser, Nucl. Phys. {\bf B195} (1982) 76.}

\REF\THE{Y. Choquet-Bruhat and J. Marsden, Commun. Math. Phys.
{\bf 51} (1976) 283; R. Shoen and S.T. Yau, Comm. Math. Phys.
{\bf 65} (1979) 45; {\bf 79} (1981) 231.}

\REF\NES{E. Witten, Comm. Math. Phys. {\bf 80} (1981) 381;
J. Nester, Phys. Lett. {\bf 83A} (1981) 241.}

\REF\MAX{G.W. Gibbons and C.M. Hull, Phys. Lett. {\bf 109B} (1982)
190;  G.W. Gibbons, G. Horowitz, S.W. Hawking and M. Perry, Comm.
Math. Phys. {\bf 88} (1983) 295.}

\REF\CALL{C. Callan, J. Harvey and A. Strominger, {\it Supersymmetric
String Solitons}, preprint EFI-91-66 (1991).}

\REF\CGR{M. Cveti\v c, S. Griffies and S.-J. Rey, Nucl. Phys.
{\bf B381} (1992) 301;  M. Cveti\v c and S. Griffies, Phys. Lett.
{\bf 285B } (1992) 27.}

\REF\DIL{G.W. Gibbons, Nucl. Phys. {\bf B207} (1982) 337;
G.W. Gibbons and K. Maeda, Nucl. Phys {\bf B298} (1988) 741;
D. Garfinkle, G. Horowitz and A. Strominger, Phys. Rev.
{\bf D43} (1991) 3140;  C.F.E. Holzhey and F. Wilczek,
Nucl. Phys. {\bf B380} (1992) 447.}

\REF\SUP{M. Cveti\v c, UPR-596-T preprint (1993), hep-th \# 93121, to
appear in the {\it Proceedings of the International Conference on
High Energy Physics}, Marseille, July 21-28, 1993 (World
Scientific 1994); UPR-0600-T  preprint (1994), hep-th \# 9402089.}

\REF\COS{R. Kallosh, A. Linde, T. Ort\'in, A. Peet and A. van
Proeyen, Phys. Rev. {\bf D46} (1992) 5278.}

\REF\GIBB{G.W. Gibbons and M. J. Perry, Nucl. Phys. {\bf B248}
(1984) 629.}

\REF\DUF{M.J. Duff, R.R. Khuri, R. Minasian, J. Rahmfeld, Nucl. Phys.
{\bf B418} (1994) 195.}

\REF\POS{G.W. Gibbons, D. Kastor, L.A.J. London, P.K. Townsend and
J. Traschen, Nucl. Phys. {\bf B416} (1994) 880.}

\REF\KAL{T. Kaluza, {\it Sitz. Preuss. Akad. Wiss.} {\bf K1}
(1921) 966; O. Klein, Z. Phys. {\bf 37} (1926) 895;
T. Appelquist and A. Chodos, Phys. Rev. {\bf D28} (1983) 772;
J. Scherk and J.H. Schwarz, Nucl. Phys. {\bf B153} (1979) 61.}

\REF\CHA{Y.M. Cho and P.G.O. Freund, Phys. Rev. {\bf D12} (1975) 1711;
Y.M. Cho and Pong Soo Jang, Phys. Rev. {\bf D12} (1975) 3789;
Y.M. Cho, J. Math. Phys. {\bf 16} (1975) 2029.}

\REF\CHB{Y.M. Cho, Phys. Rev. {\bf D35} (1987) 2628.}

\REF\POL{D. Pollard, J. Phys. {\bf A16} (1983) 565.}

\REF\GP{D.J. Gross and M.J. Perry, Nucl. Phys. {\bf B226} (1983) 29.}

\REF\SOR{R.D. Sorkin, Phys. Rev. Lett. {\bf 51} (1983) 87.}

\REF\GW{G.W. Gibbons and D.L. Wiltshire, Ann. Phys. {\bf 167} (1986)
201.}

\REF\CHC{Y.M. Cho and D.H. Park, J. Math. Phys. {\bf 31} (1990) 695.}

\REF\NA{W. Nahm, Nucl. Phys. {\bf B135} (1978) 149.}

\REF\ACB{C. Aragone and S. Deser, Phys. Lett. {\bf 86B} (1979) 161;
T.L. Curtright, Phys. Lett. {\bf 85B} (1975) 219;
F.A. Berends, J. W. van Holten, B. de Wit and P. van
Nieuwenhuizen, J. Phys. {\bf A13} (1980) 1643.}

\REF\FR{P.G.O. Freund, {\it Introduction to Supersymmetry},
(Cambridge University Press, 1986).}

\REF\FIV{E. Cremmer, Supergravity in 5 dimensions, in {\it Superspace
and Supergravity}, eds. S. W. Hawking and M. Ro\v{c}ek
(Cambridge University Press, 1981);  M. G\"{u}naydin,
G. Sierra and P.K. Townsend, Nucl. Phys.
{\bf B242} (1984) 244; {\bf B253} (1985) 573.}

\REF\SIX{P.S. Howe, G. Sierra and P.K. Townsend, Nucl. Phys. {\bf B221}
(1983) 331;  R. D'Auria, P. Fr\'e and T. Regge, Phys. Lett. {\bf 128B}
(1983) 44;  M. Awada, P.K. Townsend and G. Sierra, Class. Quant. Grav.
{\bf 2} (1985) L85;  H. Nishino and E. Sezgin,
Phys. Lett. {\bf 144B} (1984) 187.}

\REF\SEV{P.T. Townsend and P. van Nieuwenhuizen, Phys. Lett. {\bf 125B}
(1983) 41;  E. Bergshoeff, I.G. Koh and E. Sezgin, Phys. Rev. {\bf D32}
(1985) 1353; M. Pernici, K. Pilch and P. van Nieuwenhuizen, Phys. Lett.
{\bf 143B} (1984) 103.}

\REF\EIG{W. Awada and P.K. Townsend, Phys. Lett. {\bf 156B} (1985) 51.}

\REF\NIN{S.J. Gates, H. Nishino and E. Sezgin, Class. Quant. Grav.
{\bf 3} (1986) 21;  M. Awada, P.K. Townsend, M. G\"unaydin
and G. Sierra, Class. Quant. Grav. {\bf 2} (1985) 801.}

\REF\TEN{A.H Chamseddine, Nucl. Phys. {\bf B185} (1981) 403;
E. Bergshoeff, M. De Roo, B. De Wit and P. van Nieuwenhuizen,
Nucl. Phys. {\bf B195} (1982) 97;
G.F. Chapline and N.S Manton, Phys. Lett. {\bf 120B} (1983) 105.}

\REF\ELE{E. Cremmer and B. Julia, Phys. Lett. {\bf 76B} (1978) 409.}

\REF\WEA{C. Wetterich, Nucl. Phys. {\bf B211} (1983) 177;
T. Kugo and P. Townsend, Nucl. Phys. {\bf B221} (1983) 375.}

\REF\WEB{C. Wetterich, Nucl. Phys. {\bf B222} (1983) 20.}

\REF\CRE{E. Cremmer and B. Julia, Nucl. Phys. {\bf B159} (1979) 141.}

\REF\CC{Y. Choquet-Bruhat and D. Christodoulou, Acta. Math. {\bf 146}
(1981) 124; T. Parker and C.H. Taubes, Commun. Math. Phys. {\bf 84}
(1981) 124; O. Reula, J. Math. Phys. {\bf 23} (1982) 810.}

\REF\CVE{M. Cveti\v{c}, S. Griffies and S.- J. Rey, Nucl. Phys.
{\bf B389} (1993) 3.}

\REF\STW{A. Shapere, S. Trivedi and F. Wilczek, Mod. Phys. Lett.
{\bf A6} (1991) 2677.}

\REF\SEN{E. Cremmer, J. Scherk and S. Ferrara, Phys. Lett.
{\bf 74B} (1978) 61; A. Sen, Nucl. Phys. {\bf B404} (1993) 109.}

\REF\ORT{T. Ort\'in, Phys. Rev. {\bf D47} (1993) 3136.}

\REF\HAW{S. Hawking, Commun. Math. Phys. {\bf 43} (1975) 199.}

\REF\GH{G.W. Gibbons and S.W. Hawking, Phys. Rev. {\bf D15}
(1977) 2752; G.W. Gibbons and M.J. Perry, Proc. R. Soc. London
{\bf A358} (1978) 467.}

\REF\HH{J.B. Hartle and S.W. Hawking, Phys. Rev. {\bf D13} (1976) 2188.}

\REF\BEK{J.D. Bekenstein, Phys. Rev. {\bf D7} (1973) 2333; {\bf D9}
(1974) 3292; S.W. Hawking, Phys. Rev. Lett. {\bf 26} (1971) 1344;
J.M. Bardeen, B. Carter and S.W. Hawking, Comm. Math. Phys.
{\bf 31} (1973) 161; I. Moss, Phys. Rev. Lett. {\bf 69} (1992) 1852;
M. Visser, Phys. Rev. {\bf D48} (1993) 583;
R. Kallosh, T. Ort\'in and A. Peet, {\bf D47} (1993) 5400.}

\REF\STAT{J.D. Bekenstein, Phys. Rev. {\bf D7} (1973) 2333;
{\bf D12} (1975) 3077; {\bf D13} (1976) 191.}

\REF\PRES{P. Preskill, P. Schwarz, A. Shapere, S. Trivedi and
F. Wilczek, Mod. Phys. Lett. {\bf A6} (1991) 2353.}

\chap{Introduction}

An important feature of a soliton, which is defined as a
time-independent solution of classical equations of
motion with a finite energy  in a non-linear
field theory, is that it saturates the Bogomol'nyi bound for its
energy. This  bound is determined by the topological charge
for a type of  configurations.
The soliton configuration with a given topological
charge is stable against decay into
another configuration with a different topological charge. In flat
space-time the Bogomol'nyi bound\refmark{\BOG} for the energy of
the configuration  can be obtained by completing the square of
the energy density $T_{tt}$ of the configuration. The soliton satisfies
the first-order differential equations, the so-called Bogomol'nyi
or self-dual equations.

The energy
\foot{In the Einstein theory of gravity there is no intrinsic definition
of a local energy density due to the equivalence principle.
Therefore, one has to define the energy of a system
as a global quantity which is defined with respect to
background (or asymptotic) space-time.}
of a configuration in an asymptotically flat or (anti-) De Sitter
\foot{However, the formalism developed in Ref.\ADM\ is completely
general and can be applied to any type of background space-time.}
space-time is given by the ADM mass,\refmark{\ADM} which is defined
in terms of a surface integral of the conserved
current $J^{\mu} = T^{\mu\nu}K_{\nu}$
over a space-like hypersurface at spatial infinity.  Here
$T^{\mu\nu}$ is the energy-momentum tensor density and $K_{\nu}$ is
a time-like Killing vector of the asymptotic space-time.
Efforts\refmark{\THE, \NES} have been made to prove the
positivity of the ADM mass of gravitating systems, {\ie ,} the
so-called positive-energy theorems, thereby proving that the
background space-time is the lowest-energy stable state. Such proofs
involve\refmark{\NES} the evaluation of the
surface integral of the corresponding Nester's
two-form and the volume integral of its covariant divergence. Both
integrals are related through  the Stokes theorem.  The Nester's
two-form is defined in terms of a bi-linear in a spinor, which is
assumed to satisfy Witten's condition,
and a gravitational covariant derivative acting on the same  spinor.
The surface integral yields the ADM mass for the corresponding system
and the volume integral assures that the ADM mass is
a positive quantity if the matter stress-energy
tensor, if any, satisfies the dominant energy condition.

In curved space-time, the Bogomol'nyi bound for  the energy of a
soliton can be obtained by embedding the soliton solution
into supergravity.  Then, the topological charges of
soliton configurations (\eg , monopole charges for
electromagnetically charged black holes) are identified
as central charges of extended supersymmetry algebra.  For a given
set of Killing spinors, defined as spinor fields which are
constant with respect to the supercovariant derivative,
one can define a set of conserved anti-commuting supercharges.
Supersymmetric variation of the supercharge (or anti-commutation
of two supercharges) gives rise to the surface integral whose
integrand is a generalized Nester's form (with the spinor now
being the parameter of supersymmetry transformation and
the gravitational covariant derivative replaced by the supercovariant
derivative).  The surface integral gives rise to the ADM
4-momentum\refmark{\ADM} plus topological charges of
the configurations in the form which corresponds to the
anti-commutation relation of supercharges in extended supersymmetry.
The integrand in the volume integral consists of
terms, bilinear in supergravity transformations for the
fermionic fields under consideration.
This integrand is semi-positive definite provided spinors satisfy
the modified Witten's condition, and is zero if  and only if
supersymmetric variations of all fermionic
fields (under consideration) vanish. As in the  flat space-time case,
the energy of a gravitating system is then bounded from below
by the  topological charge of the configuration.
The solution that saturates the corresponding
Bogomol'nyi bound, \ie , the minimum energy configuration
for a given topological charge, satisfies the first order
differential equations (Killing spinor equations) which are
obtained by taking supersymmetric variations of fermionic
fields equal to zero.  This configuration is a
bosonic configuration which is invariant
under supersymmetry transformations, and therefore it is called
supersymmetric.
\foot{See, for example, Ref.\CALL for a general introduction on
supersymmetric solitons.}

Embedding of black hole (BH) solutions of  Maxwell-Einstein gravity
into (extended) supergravity was first done in Ref.\MAX , where it
was shown that the mass of a BH is bounded from
below by its charge.\foot{Supersymmetric embedding of another type of
topological defects, \ie , the domain wall solutions,  in $N=1$
supergravity theory was done in Ref.\CGR . The global space-time
structure\refmark{\CGR} of domain walls  bears remarkable
similarities with global space-time of the corresponding
supersymmetric charged BH's.}

Another interesting class of BH solutions, which has been subject of
intense studies,  arises in  gravity theories with
non-trivial couplings of a scalar field (``the dilaton'') to gauge
fields.  The scalar-Maxwell couplings are
common features of unified  theories, \eg , Kaluza-Klein
theories and superstring theories.  The presence of
such  scalar-Maxwell couplings
changes drastically the space-time and thermodynamic
properties of the corresponding BH solutions.
Electro-magnetically charged solutions with arbitrary  dilaton
($\varphi$) couplings $\alpha$ to the gauge kinetic term, \ie ,
${\rm e}^{\alpha \varphi}F_{\mu\nu}F^{\mu\nu}$, have been
obtained \refmark{\DIL} and their properties are shown to
depend  crucially on the value  of  the coupling $\alpha$.
For $0< \alpha <1$, the extreme charged BH's have zero Hawking
temperature and their singularities coincide with the event horizon,
\ie, an outside observer cannot  observe the singularity.  For $\alpha
= 1$, it has finite, non-zero temperature and the singularity is still
covered with the horizon.  However, for $\alpha > 1$, the
temperature becomes infinite and the singularity
becomes naked, \ie , the outside observer can see it.
\foot{Recently, it has been observed\refmark{\SUP} that
supersymmetric (extreme) domain wall
solutions in $N=1$ supergravity with a  linear supermultiplet (whose
coupling  is parameterized by the  parameter $\alpha$) exhibit
complementary features; solutions with $\alpha = 1$ separate the
solutions with the  (planar) naked singularity  and infinite
temperature ($\alpha <1$), and those with the horizon and
zero temperature ($\alpha>1$).
Note, in this paper $\alpha$ is related to the parameter $\alpha$ of
Ref.\SUP\ by taking $\alpha \to 1/\sqrt{\alpha}$.}

Supersymmetric embedding of charged  dilatonic
BH's arising in theories with an arbitrary dilaton-Maxwell coupling
$\alpha$ is incomplete;
\foot{A supersymmetric embedding of domain wall
solutions with an arbitrary dilaton coupling $\alpha$ to the
matter potential was completed in Ref.\SUP .  There,
it was also shown that the Lagrangian density for
dilaton-Maxwell-Einstein system with an arbitrary dilaton-Maxwell
coupling $\alpha$ can be obtained from $N=1$ supergravity theory
with a linear supermultiplet, whose coupling is parameterized by
$\alpha$.}
only for special values of couplings ($\alpha =
1\refmark{\COS}, \sqrt{3}\refmark{\GIBB}$)\refmark{\DUF}
the supersymmetric embeddings are known.
In fact, $N=2$ supergravity transformations of gravitino and
dilatino fields that would give the correct Bogomol'nyi bounds
for dilatonic BH solutions with an arbitrary $\alpha$
were postulated  in Ref.\POS .

On the other hand, compactification of $(4+n)$-dimensional ($(4+n)$-d)
gravity  down to 4-d, {\ie }, Kaluza-Klein (KK) theory, could
provide a natural way to obtain a 4-d gravity theory with a dilaton-Maxwell
coupling $\alpha$  which could in principle depend on the number $n$
of extra dimensions.
\foot{In KK theory the dilaton-Maxwell coupling  is given by
${\rm e}^{\alpha \varphi}F_{\mu \nu}F^{\mu \nu}$
($\alpha = \sqrt{{n+2} \over n}$), provided scalar fields
associated with the unimodular part of the internal
metric are set to be constant.
See, for example, the introduction in the second paper of  Ref.\DIL .}
In order to find the  minimum energy configurations in such a class
of theories  one has  to consider embeddings of $(4+n)$-d KK theories into
the corresponding supergravity theories (with $5 \leq D\equiv (4+n)
\leq 11$), ``KK'' supergravity theories, which is the topic of our paper.
The fact that  through such supersymmetric embeddings one might be
able to prove the Bogomol'nyi bounds for configurations with a
dilaton-Maxwell coupling $\alpha$ different from $1$ and $\sqrt 3$
was one of the motivations for the investigation presented in this
paper. However, as it turns out,
for supersymmetric embeddings of KK theories  the additional scalar
fields (the components of the unimodular part of the internal metric)
conspire with the dilaton field (the determinant of the internal
metric) in such a way that the supersymmetric charged BH solutions
of $(4+n)$-d ($n\ge 2$) KK theories are effectively those of 6-d KK theory.

KK compactifications\refmark{\KAL} of  gravity theory in
$(4+n)$-d provide a way of unifying gauge fields and gravity by
compactifying the extra dimensions in a  higher dimensional pure
gravity.  In the simplest case, one starts  from 5-d
gravity and decomposes the  metric tensor as
$$
g^{(5)}_{AB} = \left ( \matrix{g^{(5)}_{\mu \nu} & g^{(5)}_{\mu 5} \cr
g^{(5)}_{5 \nu} & \phi} \right )\ \ .\eqn\a
$$
One assumes that the fifth dimension is  curled up into a very small
circle of radius $R$ ($x_5 = x_5 + 2\pi R$, \ie ,
$M^5 \rightarrow M^4 \times S^1$)
and, therefore, it  is not experimentally measurable.
If one further assumes that the metric
components are independent of the fifth coordinate, then fields in
Eq.\a\ transform under the general coordinate transformation,
$\delta x_A = \Lambda_A (x_{\mu})$, as
$$
\delta g^{(5)}_{\mu \nu} = \partial_{\mu} \Lambda_{\nu} +
\partial_{\nu} \Lambda_{\mu} \ \ \ \ \
\delta g^{(5)}_{5 \mu} = \partial_{\mu} \Lambda_5 \ \ \ \ \
\ \ \ \ \  \delta \phi = 0\ \  .\eqn\b
$$
One notices that the transformation for $g_{5 \mu}$ is precisely the
$U(1)$ gauge variation of the Maxwell field $A_{\mu}$ and,
thereby, one can identify $g_{\mu 5} = g_{5 \mu} = \kappa A_{\mu}$.
Here $\kappa^2=8\pi/M_{pl}^2$
is the 4-dimensional gravitational constant.
In fact, in this approximation (keeping only zero modes)
the original Lagrangian density reduces to the following 4-d one:
$$
{\cal L} = -{1 \over {2 \kappa_5^2}} \sqrt{-g^{(5)}}{\cal R}^{(5)}
= \sqrt{-g} \left[ - {1 \over {2 \kappa^2}} {\cal R} - {1 \over 4}
F_{\mu \nu} F^{\mu \nu} + ... \right ]\ \ , \eqn\d
$$
where $g^{(5)}_{\mu \nu} \equiv g_{\mu \nu} + \kappa^2 A_{\mu}
A_{\nu}$, $g=\det g_{\mu\nu}$, ${\cal R}$ is the Einstein
curvature defined in terms of the 4-d metric tensor $g_{\mu\nu}$
and  $ F_{\mu \nu}=\partial_{\mu} A_{\nu} -
\partial_{\nu} A_{\mu}$ is the $U(1)$ gauge field strength.
The 5-dimensional gravitational constant $\kappa_5$ is related to
the 4-dimensional one $\kappa$ by $\kappa_5^2=\kappa^2 (2\pi R)$.

This idea was further generalized \refmark{\CHA}\ to
unify a set of non-Abelian gauge fields with  gravity.
Instead of using the zero mode expansion of the metric tensor, one
imposes a proper internal isometry\refmark{\CHA} of the
metric tensor  for space-time with dimensionality higher
than five.

The effective theory in 4-d which is of the type \d\
can be generalized
\foot{See for example  Ref.\CHB\ and references therein.}
to the case when the components of the unimodular part of the
internal metric components, as well as the dilaton, depend on the
4-d space-time coordinates.
Components associated with the internal part of a metric tensor
act as classical 4-d scalar fields.  A class of interesting
solutions  associated with such an effective 4-d KK theory
constitutes configurations with a non-trivial 4-d space-time
dependence for the dilaton, scalar fields, gauge fields,
as well as  for the 4-d space-time  metric. In particular,
spherically symmetric charged configurations correspond to
charged  BH solutions with the dilaton and other
scalar fields varying with the spatial
radial coordinate.  We shall refer to such
configurations as charged KK BH's.
We  would like to address a special class of
charged KK BH solutions, namely supersymmetric ones.
\foot{Specific solutions in a broader class, \ie , certain
non-supersymmetric configurations, were investigated in Refs.\POL$-$\CHC .}

As we have discussed above, such configurations turn out to satisfy
the so-called Killing spinor equations, and
they saturate the corresponding Bogomol'nyi bounds;
within a class of configurations they correspond to the
minimum energy configurations and are thus of special interest.
The existence of solutions which satisfy the  Killing spinor
equations implies that the original  bosonic theory
(in our case $(4+n)$-d pure gravity) can
be embedded  into  the corresponding supersymmetry  theory
and the minimum energy configurations turn out to be those which
preserve some of these supersymmetries. Thus, they are named
supersymmetric configurations.

In this paper, we address a  class of supersymmetric 4-d charged
KK BH solutions,  arising in $(4+n)$-d ($1 \le n\le 7$) KK theories.
It is an attempt to generalize the work on  supersymmetric 4-d BH's
in 5-d KK theory, pioneered by Gibbons and Perry.\refmark{\GIBB}
The class of solutions, we are studying, is obtained by choosing
the `minimal', \ie , $N=1$ or $N=2$, supersymmetric extension of
$(4+n)$-d  gravity and by assuming that the only
bosonic fields  acquiring non-zero classical values, which
depend on the 4-d space-time coordinates, are parts of
$(4+n)$-d {\it pure}  gravity.
Namely, we shall set  background  values of the other bosonic fields,
\ie ,  $(4+n)$-d gauge fields and anti-symmetric tensors,
to zero.
\foot{An example of  configurations, where other bosonic
degrees  of freedom, \ie , $(4+n)$-d gauge fields,
are turned on, corresponds to charged supersymmetric 5-d KK BH's
studied in Ref.\POS .}
In addition, we confine our attention to  static,
spherically symmetric 4-d solutions, only.
Generalizations to more general configurations, where some  or all of
the above assumptions are relaxed are subjects of further
investigation.

The paper is organized in the following way. In Chapter 2, we spell
out the minimal supersymmetric extensions of gravity theories
in $(4+n)$-d ($1 \le n \le 7$), the corresponding supersymmetry
transformations on gravitino(s) as well as the reality and chirality
conditions on spinors in $(4+n)$-d.  In Chapter 3, we describe
 dimensional reduction  of bosonic (pure gravity part) and
fermionic degrees of freedom  in  $(4+n)$-d supergravity
theories down to 4-d.  We also write down
the corresponding supersymmetry transformations in terms of the
massless degrees of freedom of the  4-d theory.  These
transformations are the starting point for obtaining Killing
spinor equations and  the corresponding Bogomol'nyi bound. In Chapter
4, we solve the Killing spinor equations
for spherically symmetric charged configurations. The Killing
spinor equations can be formally satisfied for any $n\ge 1$.  We
show that the only consistent solution is the one where only two $U(1)$
isometry factors of the internal isometry group survive,
\ie ,\ $U(1)_E \times U(1)_M$, that is to say,
the  supersymmetric BH is dyonic with electric and magnetic
charges necessarily associated with {\it different} $U(1)$ gauge factors.
In Section 4.1, we check that the constraints on
four-component Killing spinors are compatible
with the reality and chirality conditions on spinors in the original
$(4+n)$-d supergravities ($(4+n)\le 11$), and count the remaining
independent degrees of freedom for Killing spinors in 4-d.  In
Section 4.2, we derive the Bogomol'nyi bound for
$U(1) \times U(1)$-charged configurations.  In Chapter 5, we
derive the explicit solutions for supersymmetric 4-d
$U(1)_E \times U(1)_M$-dyonic KK BH's (and
Killing spinors) and discuss their thermal properties
and 4-d space-time structure.  Conclusions are given in Chapter 6.

\chap{``Minimal'' Supergravity Theories in $(4+n)$ - Dimensions}

We shall first summarize  properties of the ``minimal''
supersymmetric extensions of  pure gravity theories in $D=4+n$
($1 \le n \le 7$) space-time dimensions ($(4+n)$-d).
In particular, we are interested in supersymmetric
transformations acting on the gravitino(s), since
those are the ones which will yield the corresponding
Killing spinor equations for the supersymmetric configurations.

$D=11$  is believed to be the highest possible dimension\refmark{\NA}
for supergravity (SG) theories.  The idea being that $D \geq 12$ SG
theories compactified down to $D=4$ yield $N \geq 9$ extended
supergravity theories which contain helicity
states $\geq {5 \over 2}$.  It has been shown \refmark{\ACB}\ that
a spin $5 \over 2$ field cannot be coupled consistently
either to gravity or to simple matter
systems.  Also, for $D \geq 12$ bosonic and fermionic degrees of
freedom cannot be matched with the Lorenzian metric
signature.\refmark{\FR}  Therefore, we will restrict
our attention to $D\leq 11$ theories.  However, as we shall see
in Chapter 4, the solutions satisfying  Killing spinor equations, \ie ,
the solutions of equations  $\delta \psi^{A,i}_{\Lambda}=0$,
can be (formally) obtained for {\it any} $n\ge 1$.

In our approach we are interested in  minimal $N$ extended SG
theories, which in $D=5,6,7$ correspond to $N=2$ and in
$D=8,9,10,11$ to $N=1$ SG theories.  Such higher
dimensional supergravity theories have been
worked out by various authors; an incomplete list of references
includes Refs.\FIV$-$\ELE .  We summarize their properties,
relevant to the present work, below.

Supersymmetric extensions of pure gravity theories in
$(4+n)$-d do not only involve  an addition of
the corresponding fermionic degrees
of freedom, \ie , at least one gravitino as a gauge field of local
supersymmetry that restores the supersymmetry invariance,  but
also {\it new bosonic} degrees of freedom, \ie , gauge
fields and antisymmetric tensors, which compensate for the mismatch
in Bose-Fermi degrees of freedom.  Except for  the pure gravity
part, we will turn off all the bosonic and the fermionic fields,
\ie, except for the gravity part we set  classical values
associated with all the other bosonic fields to zero. Thus, the
bosonic part of the Lagrangian density is of the form:
$$
{\cal L} = -{1 \over {2\kappa^2}} \sqrt{-g^{(4+n)}}{\cal R}^{(4+n)}\ \ ,
\eqn\1
$$
where ${\cal R}^{(4+n)}$ is the Ricci scalar defined
in terms of a $(4+n)$-dimensional metric $g^{(4+n)}_{MN}$
and $\kappa$ is the $(n+4)$-dimensional gravitational coupling constant.

In the case when  all the  other bosonic degrees of freedom except
those of pure $(4+n)$-d  gravity are turned off, the gravitino(s)
transforms under supersymmetry as
$$
\delta \psi^{{\bf{A}},i}_{\Lambda} = D_{\Lambda}
\varepsilon^{{\bf{A}},i} = (\partial_{\Lambda} +{1 \over 4}
\Omega_{\Lambda AB}\Gamma^{AB})\varepsilon^{{\bf{A}},i}\ \ , \eqn\2
$$
where ${\bf{A}} = 1, ..., 2^{[{{n+4} \over 2}]}$
is the index for $(4+n)$-dimensional spinors and $i=1,...,N$ labels
spinors (supersymmetry parameters or gravitinos) of $N$-extended
supergravity.  As usual, $\Gamma^{AB} \equiv \Gamma^{[A}\Gamma^{B]}$,
where $\Gamma^A$'s are gamma matrices satisfying the $SO(3+n,1)$-Clifford
algebra.  The spin-connection is defined in terms of a Vielbein
$E^A_{\Lambda}$: $\Omega_{ABC} \equiv -\tilde{\Omega}_{AB,C} +
\tilde{\Omega}_{BC,A} - \tilde{\Omega}_{CA,B}$
where $\tilde{\Omega}_{AB,C} \equiv E^{\Lambda}_{[A}
E^{\Pi}_{B]}\partial_{\Pi} E_{\Lambda C}$.  Notation
$[A \ ...\  B]$ refers to antisymmetrization of the corresponding
indices.

The nature of the corresponding spinor(s)
$\varepsilon^{{\bf A},i}$ in \2\  differs in each dimension of SG
theories. It depends on the Clifford algebra satisfied by gamma
matrices for each dimension\refmark{\WEA}.  $N=1$ extensions of
$(4+n)$-dimensional Poincar\'e gravity exist if Majorana
spinors exist and the matrix ${\cal C}\gamma^a$,
where ${\cal C}$ is a charge conjugation matrix
defined below (Eqs.(2.6)$-$(2.8)), is symmetric.  These can be
satisfied in $D=4,8,9,10,11$.  In the other dimensions, where the above
conditions do not hold, one has to introduce extended
supersymmetries, $N=2$ being minimal, and spinors satisfy alternative
reality conditions, \eg , $SU(2)$ or $USp(2)$ (pseudo-) Majorana
conditions.  Properties of spinors $\varepsilon^{{\bf A}, i}$ in \2\ for
the minimal extended SG's are as follows.

For  $D = 5$,  we have two ($i=1,2$) symplectic ($USp(2)$)
four-component ({\bf A}$=1,...,4$) spinors\foot{See for example
Ref.\FIV .} defined in terms of four dimensional
two-component (${\bf a}=1,2$) Weyl spinors $(\varepsilon^i_W)$:
$$
(\varepsilon^{i}) \equiv \left ( \matrix{\varepsilon^{i{\bf a}}_W \cr
\Omega^{i}_{j} \overline{\varepsilon}^j_{W{\bf a}}} \right )\ \ ,
\eqn\aa
$$
where $\overline{\varepsilon}^i_{W\,{\bf a}}= -(\sigma^2 )^{\bf
b}_{\bf a} \varepsilon^{*i}_{W\,{\bf b}}$ and
$\Omega $ is a ($2 \times 2$) symplectic invariant matrix, \ie ,
$\Omega=i\sigma^2$.  Here, $\sigma ^2$ is the second Pauli  matrix.

For $D = 6$,  we have two  ($i=1,2$) symplectic ($USp(2)$)
eight-component ({\bf A}$=1,...,8$) Majorana spinors, related
to each other in the following way:
$$
(\varepsilon^{i})^{ *} = \Omega^{i}_{j}{\cal B}
\varepsilon^j \ \ , \eqn\bb
$$
where $\Omega=i\sigma^2$ is an antisymmetric real metric of $USp(2)$
and an invertible matrix ${\cal B}$, acting on the index ${\bf A}$
of each spinor, is defined as $\Gamma^{\mu} =
-{\cal B}^{-1}\Gamma^{\mu *}{\cal B}$ with ${\cal B}^* {\cal B} = -1$.

For $D = 7$, two ($i=1,2$) $SU(2)$  eight-component
({\bf A}$=1,...,8$) Majorana spinors are defined as
$$
(\varepsilon^{i})^* ={\cal \varepsilon}^{i}_{j}{\cal B}
\varepsilon^j \ \ , \eqn\cc
$$
where ${\cal{ \varepsilon}}^{i}_{j}$ is an $SU(2)$ invariant
antisymmetric tensor and ${\cal B}$ is defined analogously as above.

For $D = 8, 9$, there is one 16-component (${\bf A} =1,...,16$)
pseudo-Majorana spinor satisfying
$$
\overline{\varepsilon}\equiv  {\varepsilon}^\dagger \Gamma_0 =
\varepsilon^T{\cal C}\ \ ,  \eqn\dd
$$
where ${\cal C}$ is the charge conjugation matrix satisfying
${\cal C}\Gamma_{\mu}{\cal C}^{-1} = +\Gamma^T_{\mu}$.

For $D = 10$, there is  one 32-component Majorana-Weyl spinor
satisfying
$$
\overline{\varepsilon} = \varepsilon^T{\cal  C } \ \ \ \ \ \ \
\Gamma_{11} \varepsilon = \varepsilon  \eqn\ee
$$
with ${\cal C}\Gamma_{\mu}{\cal C}^{-1} = -\Gamma^T_{\mu}$.

For $D = 11$, there is one 32-component Majorana spinor satisfying
$$
\overline{\varepsilon} = \varepsilon^T{\cal C }\ \ , \eqn\ff
$$
where ${\cal C}\Gamma_{\mu}{\cal C}^{-1} = -\Gamma^T_{\mu}$.

\chap{Supersymmetric Kaluza-Klein  Compactification}

The effective Kaluza-Klein (KK) theory in  4-d  is obtained from
($4+n)$-d pure gravity by compactifying the extra $n$ spatial
coordinates on a compact manifold.  Before addressing the
compactification Ans\" atze  we spell out our notation.
General indices running over $(4+n)$-d are denoted by
upper-case letters ($A, B, ..., \Lambda , \Pi ,...$).
Lower-case letters ($a, b, ..., \lambda , \pi ,...$)
denote indices running over the 4 space-time dimensions
and lower-case letters with tilde ($\tilde{a}, \tilde{b},...,
\tilde{\lambda}, \tilde{\pi},...$) are for the $n$ extra spatial
dimensions. Latin letters ($A, B,...,a, b,...$) denote
flat-tangent space-time indices, and Greek letters
($\Lambda , \Pi ,..., \lambda , \pi ,...$)
are reserved for curved  space-time indices.
Note also  that  the 4-d  space-time coordinates ($t, \phi, r,...$)
as variables are understood as curved.  The flat Lorenz metric
of tangent space is chosen to be ($+-- \cdot \cdot \cdot -$)
with the internal coordinates all space-like.

Compactified theories in $4$-d  with the  most general KK
Ans\"atze are  obtained\refmark{\CHB}\  by imposing
the invariance of a $(4+n)$-d metric under an isometry
of the internal space.  The KK Ans\"atze for the  Vielbein
$E^{(4+n)A}_{\Lambda}$ and the corresponding
metric $g^{(4+n)}_{\Lambda \Pi}$ are of the following form:
$$
\eqalign{E^{(4+n)A}_{\Lambda} &= \left [\matrix{{\rm e}^{-{1 \over
{2\alpha}} \varphi}{\rm e}^a_{\lambda} & {\rm e}^{{1 \over {n\alpha}}
\varphi}A^{\tilde{\lambda}}_{\lambda}\Phi^{\tilde{a}}_{\tilde{\lambda}}
\cr 0 &{\rm e}^{{1 \over {n\alpha}}\varphi}
\Phi^{\tilde{a}}_{\tilde{\lambda}}}\right ] \cr
g^{(4+n)}_{\Lambda \Pi} &= \eta_{AB}E^{(4+n)A}_{\Lambda}
E^{(4+n)B}_{\Pi} =
\left [ \matrix{{\rm e}^{-{1 \over \alpha}\varphi}g_{\lambda \pi} -
{\rm e}^{{2\varphi} \over {n\alpha}}\rho_{\tilde{\lambda}
\tilde{\pi}}A^{\tilde{\lambda}}_{\lambda} A^{\tilde{\pi}}_{\pi} &
-{\rm e}^{{2\varphi} \over {n\alpha}}\rho_{\tilde{\lambda}
\tilde{\pi}}A^{\tilde{\lambda}}_{\lambda}
\cr -{\rm e}^{{2 \varphi} \over {n\alpha}}\rho_{\tilde{\lambda}
\tilde{\pi}}A^{\tilde{\pi}}_{\pi} & -{\rm e}^{{2\varphi} \over
{n\alpha}}\rho_{\tilde{\lambda} \tilde{\pi}}} \right ]\ \ ,}
\eqn\M
$$
where $\rho_{\tilde{\lambda} \tilde{\pi}} \equiv \Phi^{\tilde{a}}_
{\tilde{\lambda}}\Phi^{\tilde{a}}_{\tilde{\pi}}$ satisfies
${\hbox{det}}\rho_{\tilde{\lambda} \tilde{\pi}}=1$, \ie ,
$\rho_{\tilde{\lambda}\tilde{\pi}}$ is the unimodular part of the
internal metric $g_{\tilde{\lambda}\tilde{\pi}}$,
and $\alpha=\sqrt{{n+2}\over n}$.

The effective 4-d Lagrangian density can then be written in terms of
the above Ans\"atze, whose components depend on the internal space
coordinates as well.  One imposes ``the right invariance'' of the
$(4+n)$-d metric $g_{\Lambda \Pi}$ under the action of an isometry of
the internal space:
$$
{\cal L}_{\xi_{\tilde{\alpha}}} g_{\Lambda \Pi} = 0 \ \ ,
\ \ \ \ \ \ \ \
[\xi_{\tilde{\alpha}}, \xi_{\tilde{\beta}}] =
f^{\tilde{\gamma}}_{\tilde{\alpha} \tilde{\beta}}
\xi_{\tilde{\gamma}}\ \ , \eqn\iso
$$
where $\xi_{\tilde{\alpha}}$'s are $n$ linearly independent Killing
vectors of the internal space and ${\cal L}_{\xi_{\tilde{\alpha}}}$
is the Lie derivative in the direction of a vector $\xi_{\tilde{\alpha}}$.
The above constraints determine the following dependence of the
metric components on the internal coordinates:
$$
\partial_{\tilde{\alpha}} g_{\mu \nu} = 0 \ \ \ \ \ \
\partial_{\tilde{\alpha}} A^{\tilde{\gamma}}_{\mu} =
-f^{\tilde{\gamma}}_{\tilde{\alpha}
\tilde{\beta}}A^{\tilde{\beta}}_{\mu}  \ \ \ \ \ \
\partial_{\tilde{\alpha}}\rho_{\tilde{\beta}\tilde{\gamma}} =
f^{\tilde{\delta}}_{\tilde{\alpha}\tilde{\beta}}
\rho_{\tilde{\delta}\tilde{\gamma}} +
f^{\tilde{\delta}}_{\tilde{\alpha}\tilde{\gamma}}
\rho_{\tilde{\beta}\tilde{\delta}} \ \ .  \eqn\dep
$$
Additionally, if the isometry of the internal space is unimodular,
then the $(4+n)$-d Einstein Lagrangian density \1\ becomes
independent of the internal coordinates and after a trivial
integration over the internal coordinates the 4-d Lagrangian density
is, after setting the 4-d gravitational constant $\kappa_4$ equal
to 1, of the following form (see for example Eq.(8) of Ref.\CHB ):
$$
\eqalign{{\cal L }= -{1 \over 2}\sqrt{-g}[{\cal R} +
{\rm e}^{-\alpha\varphi}{\cal R}_K +
{1 \over 4}{\rm e}^{\alpha\varphi}
\rho_{\tilde{\alpha}\tilde{\beta}}F^{\tilde{\alpha}}_{\mu \nu}
F^{\tilde{\beta}\mu \nu}
- {1 \over 2}\partial_{\mu}\varphi \partial^{\mu}\varphi \cr
- {1 \over 4}\rho^{\tilde{\alpha}\tilde{\beta}}
\rho^{\tilde{\gamma}\tilde{\delta}}
(D_{\mu}\rho_{\tilde{\alpha}\tilde{\gamma}})
(D^{\mu}\rho_{\tilde{\beta}\tilde{\delta}})
+ \lambda (\det\rho_{\tilde{\alpha}\tilde{\beta}} - 1)]\ \ ,}
 \eqn\kal
$$
where ${\cal R}_K$ is the Ricci scalar
\foot{This term describes the self-interactions among scalar fields
and vanishes for an Abelian isometry group, \ie , if the internal
space is an $n$-torus.}
defined in terms of the unimodular part
$\rho_{\tilde{\alpha}\tilde{\beta}}$ of the internal metric,
$F^{\tilde{\alpha}}_{\mu \nu} \equiv \partial_{\mu}
A^{\tilde{\alpha}}_{\nu} - \partial_{\nu}
A^{\tilde{\alpha}}_{\mu} - gf^{\tilde{\alpha}}_{\tilde{\beta}
\tilde{\gamma}} A^{\tilde{\beta}}_{\mu} A^{\tilde{\gamma}}_{\nu}$,
where $f^{\tilde{\alpha}}_{\tilde{\beta} \tilde{\gamma}}$ is
the structure constant for the internal isometry group and $g$ is
the gauge coupling constant of the isometry group,
is the field strength of the gauge field $A^{\tilde{\alpha}}_{\mu}$,
$D_{\mu} \rho_{\tilde{\alpha}\tilde{\beta}} = \partial_{\mu}
\rho_{\tilde{\alpha}\tilde{\beta}} -
f^{\tilde{\delta}}_{\tilde{\gamma}\tilde{\beta}}A^{\tilde{\gamma}}_{\mu}
\rho_{\tilde{\alpha}\tilde{\delta}}$ is the corresponding gauge
covariant derivative and $\lambda$ is the Lagrangian multiplier.
Note that $\alpha=\sqrt{{n+2}\over n}$ specifies the coupling
constant of the dilaton $\varphi$ to the gauge fields
in the gauge field kinetic energy terms,\ie ,
${\rm e}^{\alpha \varphi}\rho_{\tilde{\alpha}
\tilde{\beta}}F^{\tilde{\alpha}}_{\mu \nu}F^{\tilde{\beta} \mu \nu}$.

With the metric Ansatz \M , we would like to find a specific
class of 4-d configurations which satisfy the Killing
spinor equations, \ie , those for which the
gravitino transformation(s) \2\ vanishes.
Given the metric Ansatz \M , whose components depend on the 4-d
space-time coordinates as well as the internal coordinates,
one can re-express the  bosonic quantities in
the gravitino transformation(s) \2\ in terms of 4-d quantities.

The next task, however, is to decompose the
$2^{[{{n+4} \over 2}]}$-component  spinors
$\varepsilon^{{\bf A},i}$ in \2\ (${\bf A}=1,...,$
$2^{[{{n+4} \over 2}]}$, $i=1,2$), as defined in  $(n+4)$-d,
in terms of $4$-component spinors
in 4-d.  The dimensional reduction \refmark{\WEB}\ of such
spinors can be accomplished by reducing the spinor with respect to a
continuous symmetry group of the compact internal space.
A $(4+n)$-dimensional spinor index ${\bf A}$ is split into
${\bf{ A}}=({\bf a},{\bf m})$ where ${\bf a} = 1,...,4$ and
${\bf m} = 1,...,2^{[{n \over 2}]}$. A $(4+n)$-d spinor then
decomposes as $\varepsilon^{\bf{A}} \equiv \varepsilon^{{\bf{(a,m)}}}$,
thus corresponding to $2^{[{n \over 2}]}$ copies of  4-component
spinors in $D = 4$. Note, that such a decomposition is valid
for each spinor, {\ie}, $i=1,2$.

For $(4+n)$-d gamma matrices $\Gamma^A$, which satisfy
the $SO(3+n,1)$ Clifford algebra $\{ \Gamma^A , \Gamma^B \} =
2\eta^{AB}$, one can always find a representation\refmark{\CHB, \WEB}
in which
$$
\Gamma^a = \gamma^a \otimes I \ \ \ \ \ \ \ \ \ \ \
\Gamma^{\tilde{a}} = \gamma^5 \otimes \gamma^{\tilde{a}}\ \ ,  \eqn\N
$$
where $\{ \gamma^a , \gamma^b \} = 2\eta^{ab}$, $\{\gamma^{\tilde{a}}
, \gamma^{\tilde{b}} \} = -2\delta^{\tilde{a} \tilde{b}}$, $I$ is the
$(2^{[{n \over 2}]} \times 2^{[{n \over 2}]})$ identity matrix and
$\gamma^5 \equiv i\gamma^0 \gamma^1 \gamma^2 \gamma^3 $.
$(\gamma^a )^{\bf{a}}_{\bf{b}}$ acts on the index ${\bf a}$ and
$(\gamma^{\tilde{a}})^{\bf m}_{\bf n}$ acts on the index ${\bf m}$
of the spinor $\varepsilon^{{\bf{(a,m)}}}$.
$\Gamma^{AB}$ can be expressed in terms
of $\gamma$'s  by applying the definition of tensor
product of matrices: $[A \otimes B]^{({\bf{a,m}})}_{({\bf{b,n}})}
\equiv A^{\bf{a}}_{\bf{b}}B^{\bf{m}}_{\bf{n}}$.
In the following we shall suppress the indices ${\bf a}=1,...,4$
(denoting components of a four-component  spinor) and keep
only the indices ${\bf m}=1,...,2^{[{n\over 2}]}$
(denoting the ${\bf m}^{th}$  four-component spinor).

With the above conventions and the assumption that the four-component
spinors depend on the 4-d space-time coordinates,
\foot{The right invariance requires spinors to be independent of
the internal coordinate.\refmark{\CHB}}
the gravitino transformation(s) \2\ reduces to the following forms:
$$
\eqalign{& \delta \psi^{\bf m}_{\mu} = \partial_{\mu}\varepsilon^{\bf m}
+ {1 \over 4} \{\omega_{\mu ab} - {1 \over \alpha}{\rm e}^c_{\mu}
\eta_{c[a}{\rm e}^{\nu}_{b]} \partial_{\nu} \varphi
+ {1 \over 2}{\rm e}^{\alpha \varphi}\rho_{\tilde{\mu}
\tilde{\lambda}}A^{\tilde{\mu}}_{\mu} F^{\tilde{\lambda}}_{ab} \}
(\gamma^{ab}) \varepsilon^{\bf m}
\cr & + {1 \over 2}{\rm e}^{{\alpha \over 2}\varphi} \sum_{\tilde{b}}
\{{1 \over 2}{\rm e}^c_{\mu}
\Phi_{\tilde{\lambda} \tilde{b}} F^{\tilde{\lambda}}_{ca} -
{1 \over {n\alpha}} \Phi_{\tilde{\mu} \tilde{b}}
A^{\tilde{\mu}}_{\mu} \partial_a \varphi -
{1 \over 2}A^{\tilde{\mu}}_{\mu} D_a
\Phi_{\tilde{\mu} \tilde{b}} - {1 \over 2}A^{\tilde{\mu}}_{\mu}
\Phi^{\tilde{c}}_{\tilde{\mu}} \Phi^{\tilde{\lambda}}_{\tilde{b}}
D_a \Phi_{\tilde{\lambda}\tilde{c}}\}(\gamma^{a5})
\tilde{\varepsilon}^{\bf m}_{\tilde{b}} \cr
& + {1 \over 2}(\Phi^{\tilde{\lambda}}_{\tilde{a}}D_{\mu}
\Phi_{\tilde{\lambda}\tilde{b}} -
\Phi^{\tilde{\lambda}}_{\tilde{b}}D_{\mu}\Phi_{\tilde{\lambda}\tilde{a}})
\tilde{\varepsilon}^{\bf m}_{\tilde{a}\tilde{b}}}
\eqn\O
$$
$$
\eqalign{\delta \psi^{\bf m}_{\tilde{\mu}}
&= {1 \over 8}{\rm e}^{\alpha \varphi}
\rho_{\tilde{\mu} \tilde{\lambda}}F^{\tilde{\lambda}}_{ab}
(\gamma^{ab})\varepsilon^{\bf m} - {1 \over 2}
{\rm e}^{{\alpha \over 2 }\varphi}
\{{1 \over {n\alpha}} \Phi_{\tilde{\mu}\tilde{b}} \partial_a \varphi
+ {1 \over 2}(D_a \Phi_{\tilde{\mu} \tilde{b}} +
\Phi^{\tilde{c}}_{\tilde{\mu}}\Phi^{\tilde{\lambda}}_{\tilde{b}} D_a
\Phi_{\tilde{\lambda} \tilde{c}})\}(\gamma^{a5})
{\tilde{\varepsilon}}^{\bf m}_{\tilde{b}} \cr
&- {1 \over 2}(f^{\tilde{\alpha}}_{\tilde{\beta}\tilde{\gamma}}
\rho_{\tilde{\alpha}\tilde{\mu}}\Phi^{\tilde{\beta}}_{\tilde{a}}
\Phi^{\tilde{\gamma}}_{\tilde{b}} -
f^{\tilde{\alpha}}_{\tilde{\beta}\tilde{\mu}}
\Phi^{\tilde{\beta}}_{\tilde{a}} \Phi_{\tilde{\alpha}\tilde{b}}
+ f^{\tilde{\alpha}}_{\tilde{\beta}\tilde{\mu}}
\Phi^{\tilde{\beta}}_{\tilde{b}} \Phi_{\tilde{\alpha}\tilde{a}})
\tilde{\varepsilon}^{\bf{m}}_{\tilde{a}\tilde{b}}\ \ ,}
\eqn\OP
$$
where $\mu=0,1,2,3$, $(a,b)=0,1,2,3$,  $\tilde\mu= 4,5,...., (3+n)$,
and $(\tilde a,\tilde b)=4,5,...,(3+n)$. The spinors with
tilde are defined as ${\tilde{\varepsilon}}_{\tilde{b}}^{\bf{m}}
\equiv(\gamma^{\tilde{b}})^{\bf{m}}_{\bf{n}}\varepsilon^{\bf{n}}$,
${\tilde{\varepsilon}}_{\tilde{a}\tilde{b}}^{\bf
m} \equiv (\gamma^{\tilde{a}\tilde{b}})^{\bf m}_{\bf n}
\varepsilon^{\bf n}$, $\Phi^{\tilde{\pi}}_{\tilde{a}} =
(\Phi^{-1})^{\tilde{a}}_{\tilde{\pi}}$
and the index $\tilde{a}$ in $\Phi^{\tilde{a}}_{\tilde{\mu}}$
is lowered by $\eta_{\tilde{a} \tilde{b}}$.
Recall, $\gamma^{ab}\equiv \gamma^{[a}\gamma^{b]}$, with $[a\ ...\ b]$
denoting antisymmetrization of the corresponding indices.

We would like to point out that the  above  formal de-composition of
the bosonic and the fermionic degrees of freedom and the corresponding
supersymmetry transformations \O\ and \OP\  can be
done for {\it any} space-time dimensions $D=4+n$.

\chap{Features of Abelian Supersymmetric Solutions}

We will confine the analysis of supersymmetric solutions to the
case of  Abelian compactifications, only.  In this case  we would
like to show that  4-d supersymmetric configurations,
which are charged, static and spherically symmetric, exist if and
only if the vacuum configurations break an Abelian isometry group $G$ of
the internal space down to $U(1)_E \times U(1)_M$, {\ie}, such
configurations correspond to 4-d dyonic black holes (BH's) whose
electric and magnetic charges are necessarily associated with
different $U(1)$ gauge factors. This constraint arises from
the fact that the Killing spinor
equations arising from \O\ and \OP\ impose consistent constraints
on the phases of spinors only  when the internal isometry group is
broken down to the $U(1)_E \times U(1)_M$ group. Namely, we shall see
that if there are more than one massless gauge
fields of the same type (electric or magnetic) or any one of gauge
fields has both electric and magnetic charges, then
one is not able to satisfy the Killing spinor equations and thus
such solutions are not supersymmetric.
We suspect that supersymmetric vacuum solutions may break perhaps
{\it any} isometry group $G$ down to  $U(1)_E \times U(1)_M$,
however, we show this explicitly in the  case for an Abelian  isometry
group, only. We shall also show that the constraints on four
component spinors $\varepsilon^{\bf m}$ are consistent with
reality and chirality conditions on the original Dirac spinors of
the underlying $(4+n)$-d supergravities as discussed at the
end of Chapter 2.  And we shall drive the Bogomol'nyi bound for
$U(1) \times U(1)$-charged black hole configurations.

If the isometry group  $G$ of the internal space is
Abelian, \ie , $U(1)^n$, then the vacuum
configurations correspond to flat internal space. The
structure constant vanishes, \ie , $f_{\alpha \beta}^\gamma=0$,
and the metric components $g_{\Lambda\Pi}$
are {\it independent} of the internal coordinates (see  Eq.\dep ).
With a proper choice of gauge, the internal metric
$\rho_{\tilde{\alpha}\tilde{\beta}}$ (see Eq.\M ) can be diagonalized:
$$
\rho_{\tilde{\alpha}\tilde{\beta}} = {\rm diag}(\rho_1 ,...,\
\rho_{n-1},\prod^{n-1}_{k=1}\rho^{-1}_k) \ \ .\eqn\P\
$$
So, indices $\tilde{a}$ and $\tilde{\alpha}$ of the fields in equations
\O\ and \OP\ take the same values, and therefore for simplicity of
notation we shall just replace the curved index $\tilde{\alpha}$
in the gauge fields by the flat index $\tilde{a}$.
The 4-d space-time metric is chosen to be of the following
spherically symmetric form:
$$
ds^2 = g_{\mu \nu}dx^{\mu} dx^{\nu} = \lambda (r)dt^2 -
\lambda^{-1}(r)dr^2 -
R(r)(d\theta^2 + \sin^2 \theta d\phi^2 ) \eqn\8
$$
and the internal metric modes  $\varphi$ and $\rho_{1,...,(n-1)}$
are functions of the radial coordinate $r$, only.

Given the spherical Ansatz for the metric,  the orthonormal
tangent frame is defined with  Vierbein components of the
following form:
$$
{\rm e}^{\hat t}_{{t}} = \lambda^{1/2} \ \ \ \ \ \
{\rm e}^{{\hat \theta}}_{{\theta}} = R^{1/2} \ \ \ \ \ \
{\rm e}^{{\hat \phi}}_{{\phi}} = R^{1/2}\sin \theta \ \ \ \ \ \
{\rm e}^{\hat r}_{{r}} = \lambda^{-1/2} \ \ , \eqn\A
$$
which yield  the metric $g_{\mu \nu} = \eta_{ab}{\rm e}^a_{\mu}
{\rm e}^b_{\nu} $ defined in  Eq.\8. Here  $\eta_{ab} =
{\rm diag}(1, -1, -1, -1)$, where $a, b ={\hat t},{\hat  \theta },
{\hat  \phi} ,{\hat  r}$  correspond to the tangent (flat)
space indices, and the flat space gamma matrices
$\gamma^{0,1,2,3}$ are ordered in the same manner,
\ie , $\gamma^{\hat{t}} = \gamma^0 ,..., \gamma^{\hat{r}} =
\gamma^3 $.

The Ans\"atze for electric and magnetic fields, compatible with
spherical symmetry, are of the following form:
$$
F_{tr}^{\tilde a}= E^{\tilde a}(r)\ \ , \ \ \ \ \ \ \ \ \
F_{\theta \phi }^{\tilde a} = P^{\tilde a} \sin\theta \ \ , \  \
{\tilde a}=4,...,(n+3)\ \ , \eqn\em
$$
where $E^{\tilde a}(r) = {\tilde{Q}^{\tilde a} \over
{R {\rm e}^{\alpha \varphi}\rho_i}}$
($i\equiv\tilde a-3=1,...,n$) is obtained from the Gauss's law  by
using the Maxwell equations $\nabla_{\mu}({\rm e}^{\alpha\varphi}\rho_i
F^{{\tilde a}\mu\nu})=0$ ($i \equiv \tilde{a} - 3=1,...,n$)
derived from the Lagrangian density (3.4).
$P^{\tilde a}$ is the physical magnetic monopole charge and the
constant  ${\tilde Q}^{\tilde a}$ is related to the physical
electric charge
\foot{We define the physical electric charge $Q$ as
$E \sim {Q \over r^2}$ in the limit $r \rightarrow \infty$.}
$Q^{\tilde a}$ of the configuration in the following way:
$Q^{\tilde a} = {\rm e}^{-\alpha \varphi_{\infty}}
\rho^{-1}_{i\infty} \tilde{Q}^{\tilde a}$. Here the subscript
$\infty$ denotes the asymptotic values of fields at infinity.

With the above Ans\" atze the Killing spinor equations, which are
obtained by setting  \OP\  equal to zero, can be cast
in the following form:
$$
\tilde{Q}^{\tilde a}R^{-1}{\rm e}^{-\alpha \varphi}
\rho^{-1}_{i} \gamma^{03}\varepsilon + P^{\tilde a}R^{-1}
\gamma^{12}\varepsilon +2{\rm e}^{-{\alpha \over 2}\varphi}
\rho^{-{1 \over 2}}_{i}\lambda^{1\over 2}
({1 \over {n \alpha}}\partial_r \varphi +
{1 \over 2}\rho^{-1}_i \partial_r \rho_i )
(\gamma^{35} \otimes \gamma^{\tilde a})\varepsilon = 0 \ \ ,
\eqn\OPP\
$$
where $i= \tilde a-3= 1,...,n$.  Recall $\rho_n=\prod_{k=1}^{n-1}
\rho_k^{-1}$ and $\alpha=\sqrt{{n+2}\over n}$.
The $n$ equations in \OPP\ impose the
following $n$ constraints between the lower two-component
spinors $\varepsilon^{\bf m}_\ell$ and the upper
two-component spinors $\varepsilon^{\bf m}_u$
($(\varepsilon^{\bf{m}})^T\equiv(\varepsilon^{\bf{m}}_u ,
\varepsilon^{\bf{m}} _\ell)^T )$):
$$
(\gamma^{\tilde a})^{\bf{m}}_{\bf{n}} \ \varepsilon^{\bf{n}}_\ell =
\eta_{\tilde{a}}{\rm e}^{i\theta_{\tilde a}(r)}
\varepsilon^{\bf{m}}_u \ ;\ \ \  \eta_{\tilde{a}}=\pm 1 \eqn\SC
$$
for each $\tilde a$ ($\tilde a= 4,...,(n+3)$) for which there are
non-zero $\tilde{Q}^{\tilde a}$ and (or) $P^{\tilde a}$.
The phase $\theta_{\tilde a}$ is defined as
$$
{\rm e}^{i \theta_{\tilde a}(r)} \equiv
\left [({{\tilde{Q}^{\tilde a} \over {{\rm e}^{\alpha \varphi}
\rho_i}} - iP^{\tilde a}})
/ ({\tilde{Q}^{\tilde a} \over {{\rm e}^{\alpha \varphi}
\rho_i}} + iP^{\tilde a}) \right ]^{1\over 2} \ \ . \eqn\phase
$$

Note that any two constraints of the type \SC\ are compatible as long
as the corresponding phase difference satisfies:
$$
\theta_{\tilde a} - \theta_{\tilde b} = \pm {\pi \over 2}\ \ ,
{\tilde a} \neq {\tilde b}. \eqn\pt
$$
Namely, $\gamma^{\tilde{a}} \varepsilon_\ell = \eta_{\tilde{a}}
{\rm e}^{i\theta_{\tilde{a}}} \varepsilon_u$ and
$\gamma^{\tilde{b}}\varepsilon_\ell = \eta_{\tilde{b}}
{\rm e}^{i\theta_{\tilde{b}}} \varepsilon_u$ imply
$\gamma^{\tilde{a}}
\gamma^{\tilde{b}} \varepsilon_\ell = \eta_{\tilde{a}}\eta_{\tilde{b}}
{\rm e}^{i(\theta_{\tilde{a}} - \theta_{\tilde{b}})}\varepsilon_\ell$.
Since $(\gamma^{\tilde{a}} \gamma^{\tilde{b}})^2 = -1$,
the constraint $\gamma^{\tilde{a}}\gamma^{\tilde{b}} \varepsilon_\ell
= \eta_{\tilde{a}}\eta_{\tilde{b}}{\rm e}^{i(\theta_{\tilde{a}}
- \theta_{\tilde{b}})}\varepsilon_\ell$
has a non-zero solution  for $\varepsilon_\ell$ if and only if \break
$\eta_{\tilde{a}} \eta_{\tilde{b}} {\rm e}^{i(\theta_{\tilde{a}} -
\theta_{\tilde{b}})} = \pm i$ and thus Eq.\pt\ has to be
satisfied.  Note, the condition \pt\ can be satisfied for {\it two}
$U(1)$ gauge factors, only.  Namely, with  more than two $U(1)$ gauge
factors one has $\theta_{\tilde a} - \theta_{\tilde b} = \pm {\pi \over 2}$
(${\tilde a} \neq {\tilde b}$) and $\theta_{\tilde b} -
\theta_{\tilde c} = \pm {\pi \over 2}$  (${\tilde  b} \neq
{\tilde c}$), which in turn  imply $\theta_{\tilde a} -
\theta_{\tilde c} = 0, \pm \pi$  (${\tilde a} \neq
{\tilde c}$). The latter constraint is thus inconsistent with
Eq.\pt .  {\it  Therefore, by  considering Killing
spinor equations, arising by setting \OP\ to zero, the allowed
supersymmetric vacua in 4-d are those with $U(1) \times U(1)$
internal isometry groups.}
Without loss of generality, in further discussion we choose two
$U(1)$ factors to be associated with the $(n-1)^{th}$ and
the $n^{th}$ internal dimensions, \ie , $\tilde a=n+2,\  n+3$.

We would now like to show, using the  Killing spinor equations
obtained by setting \O\ to zero, that the supersymmetric solution
corresponds to dyonic BH's with electric and magnetic
charges necessarily associated with different
$U(1)$ factors. Using the Ans\"atze for the dilaton
$\varphi(r)$, the internal metric (Eq.\P ), the 4-d metric
(Eq.\8), and the gauge  fields (Eq.\em ) as well as
the constraint that the Killing spinors are independent
of $t$, the Killing spinor equations arising
from Eq.\O\ are of the following form:
$$
(\lambda^{\prime} - {1 \over \alpha}\lambda \partial_r \varphi )
\gamma^{03}\varepsilon - {\rm e}^{-{\alpha \over 2}\varphi}
\lambda^{1 \over 2}R^{-1}\sum^{n+3}_{\tilde a = n+2} \rho_i^{-{1
\over 2}} \tilde{Q}^{\tilde a}(\gamma^{35} \otimes
\gamma^{\tilde a})\varepsilon = 0  \eqn\time
$$
$$
\partial_{\theta}\varepsilon - {1 \over 4} \sqrt{\lambda R}
({R^{\prime} \over R} - {1 \over \alpha}\partial_r \varphi )
\gamma^{13} \varepsilon - {1 \over 4}{\rm e}^{{\alpha \over 2}\varphi}
R^{-{1 \over 2}}\sum^{n+3}_{\tilde a = n+2} \rho^{1 \over 2}_i P^{\tilde a}
(\gamma^{25} \otimes \gamma^{\tilde a})\varepsilon = 0  \eqn\thet
$$
$$
\partial_{\phi}\varepsilon - {1 \over 2}\cos\theta \gamma^{21}
\varepsilon - {1 \over 4}\sqrt{\lambda R}\sin\theta
({R^{\prime} \over R} - {1 \over \alpha} \partial_r \varphi )
\gamma^{23} \varepsilon + {1 \over 4}{\rm e}^{{\alpha \over
2}\varphi}R^{-{1 \over 2}}\sum^{n+3}_{\tilde a = n+2}\rho^{1 \over 2}_i
P^{\tilde a}\sin\theta (\gamma^{15} \otimes
\gamma^{\tilde a})\varepsilon = 0 \eqn\ph
$$
$$
\partial_r \varepsilon + {1 \over 4}{\rm e}^{-{\alpha \over
2}\varphi}\lambda^{-{1 \over 2}}R^{-1}\sum^{n+3}_{\tilde a = n+2}
\tilde{Q}^{\tilde a}\rho_i^{-{1 \over 2}}(\gamma^{05} \otimes
\gamma^{\tilde{a}}) \varepsilon = 0 \ \ , \eqn\rad
$$
Eq.\time\ along with the spinor constraint \SC\ gives rise to the
following first order differential equation:
$$
(\lambda^{\prime} - {1 \over \alpha}\lambda\partial_r \varphi ) -
{\rm e}^{-{\alpha \over 2}\varphi}\lambda^{1 \over 2}
R^{-1}\sum^{n+3}_{\tilde a = n+2} \eta_{\tilde a}\tilde{Q}^{\tilde a}
{\rm e}^{i \theta_{\tilde a}(r)}\rho_i^{-{1 \over 2}}
= 0 \ \ .\eqn\elec
$$
The above constraint can be satisfied with one, say, the ${\tilde
a}^{th}=(n+3)^{th}$, gauge field $A^{\tilde a}_{\mu}$ having non-zero
electric charge
\foot{Note that both gauge fields $A^{n+2}_{\mu}$ and $A^{n+3}_{\mu}$
cannot have non-zero  electric (or magnetic) charges, because in
this case the phases ${\rm e}^{\theta_{({n+2},{n+3})}(r)}=\pm 1
({\rm or} \pm i)$, as determined by Eq.\phase ,
are incompatible with Eq.\pt .}
($\tilde{Q}^{\tilde a} \neq 0$), but its
magnetic charge being  necessarily zero ($P^{\tilde a}=0$). Namely,
if the field $A^{\tilde a}_{\mu}$ has
$P^{\tilde a} \ne 0$ as well,  the phase term
${\rm e}^{i\theta_{\tilde a}}$ would be complex (see Eq.\phase ),
and then Eq.\elec\ could not be satisfied with
non-zero $\tilde{Q}^{\tilde a}$.
Similarly,  a linear combination of equations \thet\ and \ph\
(after making use of Eq.(5.3), which we will derive later)
along with \SC\ yields the following differential equation:
$$
2 = \sqrt{\lambda R}({R^{\prime} \over R} - {2 \over \alpha}
\partial_r \varphi ) + i{\rm e}^{{\alpha \over 2}\varphi}R^{-{1 \over
2}}\sum^{n+3}_{\tilde a = n+2} \eta_{\tilde a}P^{\tilde a}
{\rm e}^{i\theta_{\tilde a}(r)}\rho^{1 \over 2}_i \ \ . \eqn\mag
$$
Say, if the ${\tilde a}^{th}=(n+2)^{th}$ gauge field
$A^{\tilde a}_{\mu}$ has non-zero magnetic charge $P^{\tilde a} \ne 0$,
then $A^{\tilde a}_{\mu}$ should have  {\it no} electric charge
($\tilde{Q}^{\tilde a} =0$). Namely,  in order to satisfy Eq.\mag\
the phase ${\rm e}^{i \theta_{\tilde a}(r)}$ (determined by Eq.\phase )
has to be purely imaginary, and thus it should contain {\it no}
electric  charge. In addition, the phase difference
associated with each $U(1)$ gauge factor has to be $\pm {\pi\over 2}$
(see Eq.\pt ). Thus, if one gauge field is purely electric,
the other one necessarily has to be purely magnetic.  Furthermore,
it can be shown that an alternative case where all the magnetic and
electric charges are non-zero and are chosen so that \elec\ and \mag\
are satisfied is not consistent with \pt .

{\it Therefore,  supersymmetric spherical solutions
choose the vacuum where the isometry group of the
internal space is broken down to $U(1)_E \times U(1)_M$.} These
configurations are dyonic with electric and magnetic
charges associated with {\it different} $U(1)$ factors. Recall, the
corresponding constraints between the upper and the lower
components of 4-d spinors are then of the form:
$$
\eqalign{&{\rm \ say,\ }\ {\tilde{a}}^{th}=(n+2)^{th}\
{\rm  gauge\ field\ purely\ magnetic}\
(\tilde Q^{{n+2}} = 0 ):
\ \ \gamma^{{n+2}} \varepsilon_{\ell} = i\eta_m \varepsilon_u  \cr
&{\rm \ say,\ }\ {\tilde{a}}^{th}=(n+3)^{th}\
{\rm gauge\ field\ purely\ electric}\ ({\tilde{P}}^{{n+3}} = 0 )\ \ :
\ \ \gamma^{{n+3}} \varepsilon_{\ell} =\eta_e\varepsilon_u \ \  .}
\eqn\emcons
$$
Here, $\eta_{e,m}=\pm 1$.

In the following Section, we shall explicitly check that the
constraints \emcons\ are consistent with the conditions \aa\ $-$ \ff ,
which are satisfied by the original spinors in
$D$-dimensional  supergravity (SG).

\section{Compatibility of 4-d Spinor constraints with reality and
chirality conditions on $(4+n)$-d spinors}

In this Section, we shall show that reality and chirality
conditions on Dirac spinors in each dimension $5\le D\le 11$ are
compatible with the spinor  constraints \emcons\ that we have obtained
from the Killing spinor equations, thereby, proving the existence
of Killing spinors for our BH configurations.

For $D=5$, two spinors are related through the $USp(2)$-condition
(see Eq.\aa ). Each component of spinors in equations \aa\
and \emcons\ is related by
$$
\varepsilon^{{\bf m}=i}_u = \varepsilon^i_W  \ \ \ \ \ \ \
\varepsilon^1_\ell = \overline{\varepsilon}^2_W  \ \ \ \ \ \ \
\varepsilon^2_\ell = -\overline{\varepsilon}^1_W \ \ . \eqn\five
$$
With the explicit representation $\gamma^{\tilde{4}} = i$, one can
see that the spinor constraints \emcons\ can be satisfied by a non-zero
$\varepsilon$.

For $D = 6, 7$, two spinors with each having  eight independent
components are restricted by the reality conditions \bb\ and \cc .
Without loss of generality, let's consider the case  with $i = 1$.
Since all the eight components  of the two
4-component spinors $\varepsilon^{{\bf m}=1}$ and
$\varepsilon^{{\bf{m}}=2}$, obtained  from dimensional reduction of
the 8-component spinor $\varepsilon^{i=1}$,  are independent the
constraints \emcons\ can be satisfied.  Then, it remains to show that
the second spinor $\varepsilon^{i=2} = {\cal B}^{-1}
(\varepsilon^{i=1})^*$ is a Killing spinor as well.
Since  $\varepsilon^{i=1}$ is a Killing spinor it satisfies:
$$
\hat{\nabla}_{\Pi} \varepsilon^{i=1} =
(\partial_{\Pi} + {1 \over 4}\Omega_{\Pi
AB}\Gamma^{AB})\varepsilon^{i=1} = 0 \ \ . \eqn\kill
$$
Now we perform complex conjugation of  Eq.\kill .
After making use of the definition of ${\cal B}$,
$\Gamma^{\mu *}{\cal B} = -{\cal B}\Gamma^{\mu}$ (to
turn $\Gamma^{\mu *}$ into $\Gamma^{\mu}$ and pull out the
matrix ${\cal B}$ to the left) and the fact that the matrix
${\cal B}$ is invertible, we obtain
the Killing spinor equation  \kill\ for $\varepsilon^{i=2}$ as well.

For $D = 8,...,11$, we have one Dirac spinor
constrained by a reality (and chirality) condition(s).
One may re-express the spinor constraints \emcons\ in terms of
$(4+n)$-dimensional quantities:
$$
\eqalign{&\Gamma^{\tilde a} \varepsilon = \eta_e \Gamma^0 \varepsilon
\ \ \ \ {\rm or}  \ \ \ \
\varepsilon = \eta_e \Gamma^0 \Gamma^{\tilde a}\varepsilon \ \ \ \
{\rm for\ electric\ field,\ say,}\ \ \tilde a =(n+3)\cr
&\Gamma^{\tilde b} \varepsilon = i\eta_m \Gamma^5 \Gamma^0
\varepsilon\ \ \ \ {\rm or} \ \ \ \
\varepsilon = i\eta_m \Gamma^0 \Gamma^5 \Gamma^{\tilde b} \varepsilon
\ \ \ \ {\rm for\ magnetic\ field,\ say,} \ \ \tilde b=(n+2)\ \ ,}
\eqn\high
$$
where $\Gamma^5 \equiv i\Gamma^0 \Gamma^1 \Gamma^2
\Gamma^3 $ and ${\tilde a} \ne {\tilde b}$.
There exists a basis in which all the $\Gamma$ matrices and
the spinors are real.  In this basis, the (pseudo-) Majorana
conditions \dd\ $-$ \ff\ can be written as
$$
\varepsilon = {\cal C}\Gamma^0 \varepsilon \ \ . \eqn\maj
$$
In $D=10$,  spinors are Majorana-Weyl and thus there is
additionally chirality condition:
$$
\Gamma^{11}\varepsilon = \varepsilon \ \ . \eqn\chiral
$$
Eqs.\high$-$\chiral\ are simultaneous eigen-value equations
that must be satisfied by a non-zero $\varepsilon$.  It can be shown
that the matrices $\eta_e \Gamma^0 \Gamma^{\tilde a}$, $i\eta_m \Gamma^0
\Gamma^5 \Gamma^{\tilde a}$, ${\cal C}\Gamma^0$ and
$\Gamma^{11}$ commute among themselves, by using the following
properties of the charge conjugation matrix ${\cal C}$:
$$
{\cal C} = {\cal C}^T = {\cal C}^*  \ \ \ \ \ \ \ \ \ \ \
{\cal C}\Gamma_{\Pi} = \pm \Gamma^T_{\Pi}{\cal C} \ \ , \eqn\charge
$$
the Clifford algebra $\{\Gamma^{A}, \Gamma^{B}\} = 2\eta^{AB}$ and
the following hermicity property of gamma matrices:
$$
\Gamma^{\dagger}_0 = \Gamma_0  \ \ \ \ \ \ \ \ \ \ \ \ \ \ \
\Gamma^{\dagger}_A = -\Gamma_A \ ,\ A \ne 0 \ \ .
\eqn\herm
$$
Therefore, there exists an eigen-vector $\varepsilon$ which is a
simultaneous solution of equations \high\ through \chiral .

This completes the proof of the existence of Killing spinors.  Once
one has a spinor $\varepsilon$ that satisfies the above constraints,
one substitutes it into the Killing spinor equations \OPP\ and \time\ $-$
\rad\ to obtain the explicit Killing spinor solution.

As the last point in this Section, we would like to determine the
the number of independent degrees of freedom in the Killing spinors,
which is in turn related to the number of  4-d supersymmetries which are
left unbroken by the supersymmetric configuration.
The four-component spinors $\varepsilon^{\bf m}$ in 4-d are related
to each other by the constraints \emcons\ from the Killing spinor
equations as well as the reality and chirality conditions \aa\ $-$ \ff\
on the original $2^{[{{n+4}\over 2}]}$-component Dirac spinor(s) in
$(4+n)$-d.

The Killing spinor constraints \emcons\ relate the upper and the lower
components of the four-component spinors  $\varepsilon^{\bf m}$'s, thus each
reducing the number of degrees of freedom by a factor of 2.
Reality and chirality conditions relate components of Dirac
spinors (in ($4+n$)-d) and thus they further reduce the number of
independent degrees.
A Majorana condition on Dirac spinors reduces the number of independent
degrees of freedom by another factor of 2,
\ie , a $(4+n)$-d Majorana spinor has
$2^{[{{n+4}\over 2}]}$ real independent components  while
$(4+n)$-d $SU(2)$ and $USp(2)$ (pseudo-)
Majorana spinors have $2^{[{{n+4}\over 2}]}$ complex independent
degrees of freedom.  In $D=10$, one  imposes a Majorana-Weyl
condition on a Dirac spinor; in this case the number of independent
components is further reduced by a factor of 2.

Now, for each $D=4+n$ with the minimal $N$ extended SG, one obtains
the following number of the left-over independent degrees for
the spinors and the corresponding number of 4-d
supersymmetries left intact by the supersymmetric spherical configurations:
$$
\eqalign{&{\rm  d=5\ KK\ theory:}\ \
{\rm 2\ complex\ degrees\ of\ freedom}\ (N=1)\cr
&{\rm d=6,...,10\ KK\ theories:}\
\left\{\matrix{{\rm 4\ complex\ degrees\ of\ freedom}\ (N=2),\
{\rm for}\ Q=0\ {\rm or}\ P=0 \cr
{\rm 2\ complex\ degrees\ of\ freedom}\ (N=1),\
{\rm for}\ Q \ne 0 \ne P} \right . \cr
&{\rm d=11\ KK\ theory:}\
\left\{\matrix{{\rm 8\ complex\ degrees\ of\ freedom}\ (N=4),\
{\rm for}\ Q=0\ {\rm or}\ P=0 \cr
{\rm 4\ complex\ degrees\ of\ freedom}\ (N=2),\
{\rm for}\ Q \ne 0 \ne P\ \ .} \right . }
\eqn\free
$$

\section{Bogomol'nyi Bound}
Finally, we would like to derive the Bogomol'nyi bound on the energy
of the type of BH configurations, {\ie }, electrically and
magnetically charged static, spherically symmetric  configurations,
discussed in the beginning of  Chapter 4.
For that purpose we introduce the Nester-like two-form\refmark{\NES}:
$$
\hat{E}^{\mu \nu} \equiv {1 \over 2 }\bar{\varepsilon}\Gamma^{\mu \nu
\rho}\hat{\nabla}_{\rho} \varepsilon + c.c. \ \ \ . \eqn\V
$$
Here, one has to note that $\delta \psi_{\mu 4} \equiv
\hat{\nabla}_{\rho} \varepsilon$ is the supersymmetry transformation
for the 4-d ``physical gravitino(s)''.
Namely, if the kinetic energy term for the gravitino in $(4+n)$-d
of the form
$$
{i \over 2}E \overline{\psi}_{\Gamma} \Gamma^{\Gamma \Lambda \Pi}
D_{\Lambda}\psi_{\Pi} \ \ ,  \eqn\kin
$$
where $E = \sqrt{-g^{(4+n)}}$, is expressed
in terms of the components $\psi_{\mu}^{\bf m}$  ($\mu =0,...,3$,
${\bf m}=1,...2^{[{n\over 2}]}$)
and $\psi_{\tilde{\mu}}^{\bf m}$ ($\tilde \mu =4,...,(n+3)$), as
discussed at the end of Chapter 3, then the kinetic energy term \kin\
is not diagonal in $\psi_{\mu}^{\bf m}$ and
$\psi_{\tilde{\mu}}^{\bf m}$. Therefore, one has to  perform
Weyl-rescaling and field redefinition\refmark{\CRE}  in \kin\ in order to
obtain the canonical kinetic energy terms for the physical
gravitinos and fermions in 4-d.
It turns out that the 4-d physical gravitino(s) $\psi_
{{\mu}4}^{\bf{m}}$ corresponds to the
following  combination(s) of $\psi_\mu^{\bf{m}}$ and
$\psi_{\tilde\mu}^{\bf{m}}$:
$$
\psi^{\bf m}_{\mu 4} \equiv {\rm e}^{-{1 \over
{2\alpha}}\varphi}e^a_{\mu}
[\psi^{\bf m}_{\Lambda} E^{\Lambda}_a + \gamma_a \gamma_5
(\gamma^{\tilde{a}})^{\bf m}_{\bf n} \psi^{\bf n}_{\Lambda}
E^{\Lambda}_{\tilde{a}}]\ \ . \eqn\phys
$$
Using \O\ and \OP , the supersymmetry transformation(s) on the
physical gravitino(s) (Eq.\phys ) can then be written in the
following way:
$$
\delta \psi^{\bf m}_{\mu 4} = \nabla_{\mu}\varepsilon^{\bf m} -
{i \over 4}{\rm e}^{-{\alpha \over 2}\varphi}\sum^{n+3}_{\tilde{a}=4}
\rho^{-{1 \over 2}}_i \delta^{\alpha}_{\mu}
\tilde{F}^{\tilde{a}}_{\alpha \beta} \gamma^{\beta}
(\gamma^{\tilde{a}})^{\bf m}_{\bf n}
\varepsilon^{\bf n} \ \ , \eqn\phy
$$
where $\tilde{F}^{\tilde{a}\mu \nu} \equiv {1 \over 2}e^{-1}
{\rm e}^{\alpha \varphi} \rho_i \varepsilon^{\mu \nu \alpha \beta}
F^{\tilde{a}}_{\alpha \beta}$ is the dual
electromagnetic field strength tensor and $e = \sqrt{-g}$. Recall,
$i\equiv ({\tilde a}-3)=1,..., n$,
$\rho_n = \prod_{k=1}^{n-1}\rho_k^{-1}$ and $\alpha=\sqrt{{n+2}\over n}$.
In addition, one redefines the fermionic fields as follows:
$$
\chi^{\bf m}_{\tilde{a}} \equiv {\rm e}^{-{\varphi \over {2\alpha}}}
\psi^{\bf m}_{\Lambda} E^{\Lambda}_{\tilde{a}} \ \ , \eqn\ferm
$$
where $\tilde a =4,...,(n+3)$. Using again equations \O\ and \OP , one
obtains the corresponding supersymmetry transformations given by
$$
\delta \chi^{\bf m}_{\tilde{a}} = {1 \over 8}
{\rm e}^{{\alpha \over 2}\varphi}\rho^{1 \over 2}_i
F^{\tilde{a}}_{\mu \nu} \gamma^{\mu \nu}
\varepsilon^{\bf m} + {1 \over {2n\alpha}}\partial_{\mu}\varphi
\gamma^{\mu} \gamma^5 (\gamma^{\tilde{a}})^{\bf m}_{\bf n}
\varepsilon^{\bf n} + {1 \over 4}\rho^{-1}_i
\partial_{\mu}\rho_i
\gamma^{\mu} \gamma^5 (\gamma^{\tilde{a}})^{\bf m}_{\bf n}
\varepsilon^{\bf n} \ \  . \eqn\susy
$$
In terms of the physical gravitinos \phys\ and the redefined
scalar fields \ferm , the kinetic energy  term \kin\ assumes
the following canonical form:
$$
{i \over 2}e \overline{\psi}^{\bf m}_{\mu} \gamma^{\mu \nu \rho}D_{\nu}
\psi^{\bf m}_{\rho} + {i \over 2} e \overline{\chi}^{\bf m}_{\tilde{a}}
({1 \over 2}(\gamma^{\tilde{a}}\gamma^{\tilde{b}})^{\bf m}_{\bf n} +
\eta^{\tilde{a}\tilde{b}}\delta^{\bf{m}}_{\bf{n}})\gamma^{\mu}D_{\mu}
\chi^{\bf n}_{\tilde{b}} \ \ , \eqn\kk
$$
where $D_{\mu} \equiv \partial_{\mu} + {1 \over 4}\omega_{\mu
ab}\gamma^{ab}$ is the 4-d gravitational covariant
derivative on fermionic fields.

Derivation of the Bogomol'nyi bound consists of evaluating the
surface integral of the Nester's two-form \V, which is related through the
Stokes theorem to the volume integral of its covariant derivative
in the following way:
$$
\int_{\Sigma}dS_{\mu} e \nabla_{\nu} \hat{E}^{\mu \nu} =
\int_{\Sigma}dS_{\mu} \partial_{\nu}(e \hat{E}^{\mu \nu}) =
{1 \over 2}\int_{\partial \Sigma} dS_{\mu \nu}e
\hat{E}^{\mu \nu}\ \ , \eqn\W
$$
where $\Sigma$ is a space-like hypersurface with the boundary
$\partial \Sigma$ at spatial infinity.
With the given supersymmetry transformation(s) \phy\ of
the physical 4-d gravitino(s), the Nester two-form \V\
reduces to the following expression:
$$
\hat{E}^{\mu \nu} = {1 \over 2}[\overline{\varepsilon}
\gamma^{\mu \nu \rho}
\nabla_{\rho}\varepsilon + i\overline{\varepsilon}
\{{\rm e}^{-{\alpha \over 2}\varphi} \sum^{n+3}_{\tilde{a}=4}
\rho^{-{1 \over 2}}_i(\tilde{F}^{\tilde{a}\mu \nu} + {1 \over 2}
\gamma^{\mu \nu \alpha \beta}\tilde{F}^{\tilde{a}}_{\alpha \beta})
\otimes \gamma^{\tilde{a}}\}\varepsilon ]
+ c.c. \ \ . \eqn\nest
$$

As will be shown in the next chapter,
space-time of the configuration is asymptotically flat and the spinor
$\varepsilon$ approaches  a constant, $\varepsilon_\infty$,
as $r \rightarrow  \infty$. So, the surface integral in \W\ is
evaluated to be
$$
\int_{\partial \Sigma}dS_{\mu \nu} e \hat{E}^{\mu \nu} =
(\bar{\varepsilon}_{\infty} \gamma^{\mu} \varepsilon_{\infty})
P^{ADM}_\mu +\sum^{n+3}_{\tilde{a}=4} \overline{\varepsilon}_{\infty}
{\rm e}^{{\alpha \over 2}\varphi_{\infty}}
\rho^{1 \over 2}_{i\infty}(iP^{\tilde{a}}
- \gamma_5 Q^{\tilde{a}}) \otimes \gamma^{\tilde{a}}
\varepsilon_{\infty}\ \ ,
\eqn\X
$$
where $P^{ADM}_{\mu}$ is the ADM 4-momentum\refmark{\ADM}
and the physical charges of the system are defined as
$Q^{\tilde {a}} \equiv {1 \over 2} \int_{\partial
\Sigma}dS_{\mu \nu}F^{\tilde{a}\mu \nu}$ and  $P^{\tilde {a}} \equiv
{1 \over 2}\int_{\partial \Sigma}
dS_{\mu \nu}e({\rm e}^{-\alpha\varphi} \rho^{-1}_i
\tilde{F}^{\tilde{a}\mu \nu})$. The
the subscript $\infty$ denotes the asymptotic value of a field as
$r\to \infty$.

The volume integral in Eq.\W\ is more involved and
a lengthy calculation yields
$$
\eqalign{\int_{\Sigma} dS_{\mu} e \nabla_{\nu}
\hat{E}^{\mu \nu} = &\int_{\Sigma}dS_{\mu}
e[-{1 \over 2}(\overline{\hat{\nabla}_{\nu}}\varepsilon^{\bf m})
\gamma^{\mu \nu \rho} \hat{\nabla}_{\rho} \varepsilon^{\bf m} +
{1 \over 2}(\overline{\delta \chi^{\bf m}_{\tilde{a}}})[{1 \over 2}
(\gamma^{\tilde{a}}\gamma^{\tilde{b}})^{\bf m}_{\bf n}
+ \eta^{\tilde{a}\tilde{b}}\delta^{\bf m}_{\bf n}]
\gamma^{\mu}(\delta \chi^{\bf n}_{\tilde{b}}) \cr
&+ (G^{\mu \nu} - T^{\mu \nu})(\overline{\varepsilon^{\bf m}}
\gamma_{\nu}\varepsilon^{\bf m})+ c.c.] \ \ ,} \eqn\Y
$$
where $T^{\mu \nu} = {1 \over \sqrt{-g}}{{\partial {\cal L}_{matt}}
\over {\partial g_{\mu \nu}}}$ is the stress-energy tensor for matter
(gauge fields, the dilaton and scalar fields) terms.  The first term in
the integrand on the right hand side (RHS) of Eq.\Y\ is non-negative for
spinors $\varepsilon$ satisfying the (modified) Witten's condition,
\ie , $n \cdot \hat{\nabla}\varepsilon = 0$ ($n$ is the 4-vector
normal to $\Sigma$), which is shown \refmark{\CC}\ to have a solution
for an asymptotically flat space-time provided the spinor approaches
a constant value as $r \rightarrow \infty$.  The last term on
the RHS of Eq.\Y\  vanishes due to the Einstein equation
$G^{\mu \nu} = T^{\mu \nu}$.
Therefore, \X\ is non-negative and vanishes, provided  the
supersymmetry transformations \phy\ and
\susy\ of the physical 4-d gravitinos $\psi^{\bf m}_{\mu 4}$ and the
fermions $\chi^{\bf m}_{\tilde{a}}$ are zero. Note that
vanishing of \phy\ and \susy\ is equivalent to  vanishing of
the supersymmetry transformations \O\ and \OP\ of the decomposed
$(n+4)$-d gravitino $\psi_{\Pi}$, \ie , $\delta \psi_{\mu}^{\bf m} =\delta
\psi_{\tilde{\mu}}^{\bf m}=0$.

For non-supersymmetric configurations, \Y\ implies that the bilinear form
\X\ is positive.  The necessary and sufficient condition for \X\ to
be positive is that all the eigenvalues of the Hermitian matrix
sandwiched between the spinor $\varepsilon_{\infty}$ are positive.
Since the matrix in the first term on the RHS of \X\ commutes with
the matrix in the second term (this can be easily seen by going into the
reference frame of the configuration, \ie , the frame where the ADM
energy-momentum $P^{ADM}_{\mu}$ has only time-component), the
ADM mass of the configuration, \ie , the eigenvalue of the matrix in
the first term, has to be greater than the largest eigenvalue of the
matrix in the second term.
\foot{One of ways of determining the largest eigenvalue of the matrix
in the second term is to consider all the possible sets of commuting
matrices in the second term and to express the matrix in the second
term in a form of sum of mutually anticommuting matrices
whose eigenvalues are known.  Then, the largest eigenvalue can be
found by applying the following theorem:
if matrices $A_i$ have eigenvalues $a_i$ and $A_i$'s are mutually
anticommuting, then the matrix $\sum^m_{i=1}A_i$ has
eigenvalues $\pm \sqrt{\sum^m_{i=1}(a_i)^2}$.  This method can
be applied to the case of even number of internal dimensions, only.}
We have seen that the supersymmetric configurations prefer
the vacua with two $U(1)$ gauge factors.  It is therefore of interest
for us to consider only two non-zero gauge fields, which are
without loss of generality taken to be associated with the last two
internal coordinates.  Then, applying the prescription stated in
the footnote, we obtain the following Bogomol'nyi bound for
non-supersymmetric $U(1) \times U(1)$ configurations:
$$
M > {\rm e}^{{\alpha \over 2}\varphi_{\infty}}
\sqrt{(\rho^{1\over 2}_{(n-1)\infty}|P^{n+2}| +
\rho^{1\over 2}_{n\infty}|Q^{n+3}|)^2 +
(\rho^{1\over 2}_{(n-1)\infty}|Q^{n+2}| +
\rho^{1\over 2}_{n\infty}|P^{n+3}|)^2} \ \ \eqn\bound
$$
This bound reduces to the one of Gibbons and Perry's 5-d KK BH solutions
in the limit that either of gauge fields vanishes, say,
$P^{n+2} = Q^{n+2} = 0$.

However, for supersymmetric configurations
Eq.\X\ has meaning only when the isometry group of the
internal space is $U(1)_E \times U(1)_M$. Otherwise, the constraints on
the spinor $\varepsilon_{\infty}$ cannot be satisfied, as discussed in the
beginning of this Chapter.  Say, for the gauge field  associated with
the $(n+2)^{th}$-dimension being magnetic
and the  gauge field associated with the $(n+3)^{th}$-dimension being
electric, the spinor constraints  $\gamma^{n+2} \varepsilon_{\ell\infty}
= i\eta_m \varepsilon_{u \infty}$ and $\gamma^{n+3}
\varepsilon_{\ell \infty} = \eta_e \varepsilon_{u \infty}$
($\eta_{e,m}=\pm 1$) select out magnetic charge
$P^{n+2}$ from the second to the last term and electric charge
$Q^{n+3}$ from the last term in the second term of Eq.\X .
Therefore, the ADM mass of the supersymmetric configuration becomes:
$$
M_{ext} = {\rm e}^{{\alpha \over 2}\varphi_{\infty}}
(\rho^{1\over 2}_{n\infty}|P^{n+2}|+\rho^{1\over
2}_{(n-1)\infty}|Q^{n+3}|)\ \ .\eqn\adm
$$
This is the right expression for the ADM mass that saturates the bound
\bound\ in the case of the $U(1)_E \times U(1)_M$ group.

\chap{Supersymmetric 4-d  Kaluza-Klein Solutions}

We shall now obtain the explicit form of the supersymmetric  4-d
charged  Kaluza-Klein (KK) black hole (BH) solutions satisfying
the Killing spinor equations as specified
by vanishing of the supersymmetry transformations \O\ and \OP .

We have shown in Chapter 4 that the maximal symmetry of the internal
space allowed by supersymmetric static spherical configurations is
$U(1)_E \times U(1)_M$.  Without loss of generality, we choose the
electromagnetic vector potential associated with the second to the
last coordinate to be magnetic and that corresponding to the last
coordinate to be electric:
$$
A^{n+2}_{\mu} = \delta^{\varphi}_{\mu} P(1 - \cos\theta) \ \ \ \
\ \ \ \ \ \ \ \  A^{n+3}_{\mu} = \delta^t_{\mu} \psi (r) \ \ ,
\eqn\pot
$$
where $E(r) = -\partial_r \psi (r) = {\tilde{Q} \over
{R{\rm e}^{\alpha\varphi}\rho_n}}$.

The aim is now to obtain the explicit solutions for the
4-d metric components (Eq.\8), the internal radii (Eq.\P )
and the dilaton $\varphi(r)$ from the
Killing spinor equations \OPP\ and \time\ $-$ \rad. However, before
obtaining the first order coupled differential equations for  these
fields, we have to determine the angular coordinate dependence of
the spinors $\varepsilon^{\bf m}$.  For this purpose, we multiply
\thet\ by $\gamma^1 \sin\theta$ and \ph\ by $\gamma^2$.  Then we
subtract the two and then multiply by $\gamma^2$ to get the equation
$$
[2\partial_{\phi} + \gamma^1 \gamma^2 \cos\theta -
2(\gamma^1 \gamma^2 \sin\theta) \partial_{\theta}]
\varepsilon^{\bf m}_u = 0 \ \ .
\eqn\ang
$$
This fixes the angular coordinate dependence\refmark{\CVE}
of the spinor to be
$$
(\varepsilon^{1, {\bf m}}_u , \varepsilon^{2, {\bf m}}_u ) =
{\rm e}^{i\sigma^{2}\theta / 2}{\rm e}^{i\sigma^3 \phi / 2}
(a^{1, {\bf m}}_u (r) , a^{2, {\bf m}}_u (r)) \ \ , \eqn\spin
$$
where $a^{\bf m}_u (r)$'s are two-component spinors which depend on the
radial coordinate, only. The same relation holds for the lower
two-component spinors $\varepsilon^{\bf m}_\ell$.  Constraints
on $a^{\bf m}_{u, \ell}(r)$ are the same
as those of $\varepsilon^{\bf m}_{u, \ell}$, \ie , Eq.\emcons\ with
$\varepsilon_{u, \ell}$ replaced by $a_{u,\ell}$.

Given the Ans\"atze for the gauge fields \pot , and the
constraints \emcons\ and \spin\ on spinors, one can
solve the Killing spinor equations (equations \OPP\ and \time\
through \rad ) to get the following differential equations for
the 4-d metric coefficients $\lambda(r)$ and $R(r)$ in equation \8,
the internal radii $\rho_i(r)$ ($i=1,...,n$) and the
dilaton $\varphi(r)$ as well as the spinors $a^{\bf m}_{u}(r)$:
$$
\lambda^{\prime} - {1 \over \alpha}\lambda \partial_r \varphi -
\eta_e \tilde{Q}{\rm e}^{-{\alpha \over 2}\varphi}\rho^{-{1\over 2}}_n
\lambda^{1\over 2} R^{-1}  = 0  \eqn\da
$$
$$
\sqrt{\lambda R}({R^{\prime} \over R} - {1 \over \alpha}
\partial_r \varphi ) - \eta_m P{\rm e}^{{\alpha \over 2}\varphi}
\rho^{1\over 2}_{n-1}R^{-{1\over 2}}=2  \eqn\db
$$
$$
\partial_r a^{\bf m}_u - {1\over 4}\eta_e \tilde {Q}
{\rm e}^{-{\alpha \over 2}\varphi} \rho^{-{1\over 2}}_n
\lambda^{-{1 \over 2}}R^{-1} a^{\bf m}_u = 0  \eqn\dc
$$
$$
{1 \over {n\alpha}} \partial_r \varphi + {1\over 2}\rho^{-1}_i
\partial_r \rho_i = 0 \ \ ,\ \ i = 1,...,n-2 \eqn\dd
$$
$$
PR^{-1} + 2\eta_m {\rm e}^{-{\alpha \over 2}\varphi}
\rho^{-{1\over 2}}_{n-1}\lambda^{1 \over 2} ({1\over {n\alpha}}
\partial_r \varphi + {1\over 2}\rho^{-1}_{n-1} \partial_r
\rho_{n-1} ) = 0 \eqn\de
$$
$$
\tilde{Q}R^{-1}{\rm e}^{-\alpha \varphi}\rho^{-1}_n  +
2\eta_e {\rm e}^{-{\alpha \over 2}\varphi} \rho^{-{1\over 2}}_n
\lambda^{1 \over 2}({1 \over {n\alpha}}\partial_r \varphi + {1\over
2}\rho^{-1}_n \partial_r \rho_n ) = 0 \ \ .  \eqn\df
$$
Recall, $\alpha=\sqrt{ {n+2}\over n}$,
$\rho_n=\prod_{k=1}^{n-1}\rho_k^{-1}$ and $\eta_{e,m}=\pm 1$.

We shall now solve these equations to obtain the supersymmetric
solutions for charged static spherical configurations. The 4-d metric
components $\lambda$ and $R$  are related by the following
equation:
$$
\partial_r \sqrt{\lambda R} = 1\ \ , \eqn\met
$$
which can be solved to yield
$$
\lambda R = (r - r_H )^2 \ \ , \eqn\ext
$$
where $r_H$ is the event horizon, \ie \ $\lambda (r_H ) = 0$.
Eq.\dd\ is integrated to yield the expressions
for $\rho_i$ ($i = 1,...,n-2$) in terms of $\varphi$:
$$
\rho_i = \rho_{i\infty}{\rm e}^{-{2 \over {n\alpha}}
(\varphi -\varphi_{\infty})} \ \ ,\ \  i = 1,...,n-2 \ \ ,
\eqn\reli
$$
and equations \db\ and \de\ with \ext\ are solved to give the
following relation of $\rho_{n-1}$ to $\lambda$ and $\varphi$:
$$
\rho_{n-1} = \rho_{(n-1)\infty} \lambda {\rm e}^{-{{2-n}\over
{n\alpha}}(\varphi - \varphi_{\infty})} \ \ ,
\eqn\reln
$$
where the subscript $\infty$ denotes the asymptotic values of fields
at infinity.

Making use of the relations \ext\ through \reln\ among the fields
associated with the internal metric and the 4-d metric  components,
we can rewrite the equations \da\ and \db\
entirely in terms of the 4-d  metric components and the dilaton field
$\varphi$:
$$
{\lambda^{\prime}\over \lambda} - {1 \over \alpha}\partial_r \varphi
-\eta_e {\rm e}^{-{1 \over \alpha}(\varphi - \varphi_{\infty})}
{{\bf Q}\over R} = 0  \eqn\first
$$
$$
{\lambda^{\prime} \over \lambda} + {1 \over \alpha}\partial_r \varphi
+ \eta_m {\rm e}^{{1\over \alpha}(\varphi - \varphi_{\infty})}
{{{\bf P}} \over R} = 0 \  \ ,
\eqn\secon
$$
where we have defined the following ``screened'' electric and
magnetic charges:
$$
\eqalign{{\bf Q}&\equiv {\rm e}^{{\alpha \over
2}\varphi_{\infty}}\rho^{1 \over 2}_{n\infty}Q\cr
{\bf P}&\equiv {\rm e}^{{\alpha \over 2}\varphi_{\infty}} \rho^{1
\over 2}_{(n-1)\infty}P\ \ .}  \eqn\qpbf $$
Here, $Q$ and  $P$  are the respective physical electric and magnetic
charges (see comments after Eq.\em , where the relationship between
$\tilde Q$ and the physical electric charge $Q$ is discussed).  Note
that the ADM mass of the  extreme configuration (Eq.\adm )  also
depends only on the screened charges ${\bf  Q}$ and ${\bf P}$.
In addition, notice the symmetry of the above
two equations under the electro-magnetic duality transformations, \ie ,
${\bf P}  \leftrightarrow {\bf Q}$ and $\varphi \rightarrow -\varphi$.
Subtracting the above two equations, we obtain the following equation
$$
{1 \over \alpha}\partial_r \varphi + {1\over 2}\eta_m
{\rm e}^{{1\over \alpha}(\varphi - \varphi_{\infty})}
{{\bf P} \over R} + {1\over 2}\eta_e {\rm e}^{-{1\over \alpha}
(\varphi - \varphi_{\infty})}
{{\bf Q} \over R} = 0 \ \ , \eqn\sym
$$
which confirms a {\it no-hair} theorem, \ie , the constant
dilaton field with zero electromagnetic fields ($P=Q=0$).

Multiplying Eq.\first\ by $\eta_m {\bf P}{\rm e}^{{1\over \alpha}
(\varphi - \varphi_{\infty})}$ and Eq.\secon\ by $\eta_e
{\bf Q}{\rm e}^{-{1\over \alpha}(\varphi - \varphi_{\infty})}$,
followed by addition of the two  equations,
gives the equation which is solved to be
$$
\lambda = {{\eta_e {\bf Q}{\rm e}^{-{1 \over \alpha}(\varphi
-\varphi_{\infty})} +
\eta_m {\bf P}{\rm e}^{{1\over \alpha}(\varphi - \varphi_{\infty})}}
\over {\eta_e {\bf Q}+ \eta_m {\bf P}}} \ \ .
\eqn\comp
$$
This expression relates the 4-d metric coefficient $\lambda$  and
the value of the dilaton $\varphi$, and it reduces to the following
special relations:
$$
\varphi = -\alpha \ln\lambda + \varphi_{\infty} \ \ \ \ {\rm and} \ \ \ \
\varphi = \alpha \ln\lambda + \varphi_{\infty}\ \ ,  \eqn\old
$$
which correspond to the purely  electrically charged ($P=0$) and
the purely magnetically charged ($Q=0$)
BH's, respectively.  We substitute \comp\ into \sym , making use
of \ext , in order  to get the following ordinary differential
equation for the dilaton field $\varphi$:
$$
{{1 \over \alpha}\partial_r \varphi + {1 \over {2(r - r_H)^2 }}
\left [\eta_e {\bf Q}{\rm e}^{-{1 \over \alpha}
(\varphi - \varphi_{\infty})} + \eta_m {\bf P}
{\rm e}^{{1\over \alpha}(\varphi -\varphi_{\infty})}\right ]^2
{1 \over {\eta_e {\bf Q} + \eta_m {\bf P}}} = 0 \ \ .}
\eqn\dileq
$$
Note once again the symmetry of \dileq\ under the electro-magnetic
duality transformation. This equation can be easily solved
to give the explicit solution for the dilaton field:
$$
{\rm e}^{{2 \over \alpha}(\varphi - \varphi_{\infty})} =
{{r - r_H + \eta_e {\bf Q}} \over {r - r_H - \eta_m
{\bf P}}} = {{r - |{\bf P}|} \over {r - |{\bf Q}|}}  \eqn\sol
$$
where we have identified $r_H = \eta_e {\bf Q} - \eta_m
{\bf P}$.  Also, we have chosen the signs of
$\eta_m$ and $\eta_e$ ($\eta_{e,m}=\pm 1$) so that $\eta_e{\bf Q} =
|{\bf Q}|$ and $-\eta_m {\bf P} = |{\bf P}|$.
Then, we substitute \sol\ into \comp\ to
obtain the explicit solution for $\lambda$:
$$
\lambda = {{r -|{\bf Q}|-|{\bf P}|} \over
{(r - |{\bf Q}|)^{1 \over 2}(r - |{\bf P}|)^{1 \over 2}}}\ \ ,
\eqn\met
$$
and by using \ext\ the following solution for $R$ is obtained:
$$
{R = r^2 (1 - {{|{\bf Q}| +|{\bf P}|}\over r})
(1 - {{|{\bf Q}|} \over r})^{1 \over 2}
(1 - {{|{\bf P}|} \over r})^{1 \over 2}\ \ .}
\eqn\are
$$

The following solution for the electric field:
$$
E(r) = {Q \over {R{\rm e}^{\alpha
(\varphi - \varphi_{\infty})} (\rho_n / \rho_{n\infty})}} =
{Q \over {(r - |{\bf P}|)^2 }}  \eqn\elec
$$
is obtained by substituting the explicit
solutions for $R$, $\varphi$ and $\rho_n$ into the formula derived
from the Euler-Lagrangian equation for the electric field.
The electric field has different radial dependence from that of
the axionic dyon solutions:
its radial dependence is shifted by $|{\bf P}|$.

Finally, the  radial coordinate dependence of the 4-d spinors is
fixed by equation \dc\ with the known solutions \reli , \reln\ and \sol :
$$
a^{\bf m}_{u}(r) = a^{\bf m}_{u\infty}
\left ( {{r - |{\bf Q}|-|{\bf P}|} \over {r -
|{\bf P}|}} \right )^{1 \over 4} \ \ . \eqn\sub
$$
Note that the spinors $\varepsilon^{\bf m}$ do asymptotically
approach constant spinors
$\varepsilon^{\bf m}_{u \infty}$ as $r \rightarrow \infty$.
For $Q=0$, they are  independent of the radial coordinate.
Otherwise, they approach  zero  at the  horizon.

Although the solution \sol\  for the dilaton field $\varphi$
depends explicitly on $n$ (the number of internal
dimensions) the 4-d metric coefficients
\met\ and \are\  as well as the Killing spinors, determined by
equations \spin\ and  \sub , do not.
{\it Supersymmetry, by means of the Killing spinor
equations, renders the scalar fields $\rho_i$ to collaborate with the
dilaton field $\varphi$, through \reli\ and \reln , in
such a way that the 4-d space-time properties of the configuration
are independent of dimensionality of the internal space}.
In addition, in the limit that either $Q$ or $P$ vanishes,
solutions for the 4-d metric components $\lambda$
and $R$ reduce to those of the 4-d supersymmetric BH's in  5-d KK
theory.\refmark{\GIBB}

The above properties of solutions \met\ and \are\
can be understood by calculating the effective on-shell
action (this we mean by considering the equations for
the scalar fields only) for supersymmetric configurations, \ie , by
considering the Lagrangian of the bosonic fields associated with this
class of configurations.  From equations \reli\ and \reln ,
we can see that the following combinations of the scalar fields and
$\lambda$ are constants:
$$
\eqalign{&\ln\rho_i + {2\over {n\alpha}}\varphi =
const.\ \ \ \ \ \ \ \ \ \ i=1,...,n-2 \cr
&\ln(\rho_{n-1} \lambda^{-1})+{{2-n}\over {n\alpha}}\varphi = const.
\ \ \ \ \ \  \ln(\rho_n \lambda) + {{2-n}\over {n\alpha}}\varphi =
const. \ \ ,}\eqn\const
$$
where we have to keep in mind that $\rho_n \equiv
\prod_{k=1}^{n-1}\rho_k^{-1}$. The above relations  provide us with a
hint that it is convenient to introduce new scalar fields $\chi_i$
and $\Phi$, which  represent the same physical degrees
of freedom as the old set $\varphi$ and $\rho_i$. The new fields are
defined in the following way:
$$
\eqalign{&\Phi \equiv {\sqrt{2} \over \alpha}\varphi \ \ \ \ \ \ \ \ \ \
\chi_i \equiv {1 \over \sqrt{2}}[\ln\rho_i +
{2\over {n\alpha}}\varphi]\ ,\ \ i=1,...,n-2 \cr
&\chi_{n-1} \equiv {1 \over \sqrt{2}}[\ln\rho_{n-1} +
{{2-n}\over {n\alpha}}\varphi ]  \ \ \ \ \ \ \
\chi_n \equiv {1\over \sqrt{2}}[\ln\rho_n +
{{2-n}\over {n\alpha}}\varphi]\ \ ,}
\eqn\new
$$
Note $\sum^n_{i=1}\chi_i = 0$.  Then, the Lagrangian density
\kal\  expressed in terms of the above new fields  becomes
$$
\eqalign{{\tilde {\cal L}} =-{1\over 2}\sqrt{-g}&[{\cal R} -
{1\over 2} \partial_{\mu}\Phi \partial^{\mu}\Phi  -
{1\over 2}\sum_{i=1}^{n} \partial_{\mu}\chi_{i}\partial^{\mu}\chi_{i}
-\partial_{\mu}\Phi (\partial^{\mu}\chi_{n-1}+\partial^{\mu}\chi_{n})
\cr
&+ {1\over 4}{\rm e}^{\sqrt{2}(\Phi + \chi_{n-1})}
F^{n+2}_{\mu\nu}F^{n+2 \ \mu\nu} +
{1\over 4}{\rm e}^{\sqrt{2}(\Phi +\chi_n)}
F^{n+3}_{\mu\nu} F^{n+3 \ \mu\nu}]\ \  . }
\eqn\effec
$$
Since the fields $\chi_i$ ($i=1,...,(n-2)$) couple in the above
Lagrangian density only through the 4-d metric,
the solutions to the Euler-Lagrangian equations of these fields are
$\chi_i=\chi_{i\infty}=const.$ ($i=1,...,(n-2)$), which in turn
implies $\chi_{n-1}+\chi_n=\chi_{(n-1)\infty}+\chi_{n\infty}=const.$.
The left-over part of the on-shell Lagrangian density is then of the
form:
$$
\eqalign{{\tilde {\cal L}'} =-{1\over 2}\sqrt{-g}&[{\cal R} -
{1\over 2} \partial_{\mu}\Phi \partial^{\mu}\Phi  -
{1\over 2} \partial_{\mu}\chi_{n-1}\partial^{\mu}\chi_{n-1} -
{1\over 2} \partial_{\mu}\chi_n \partial^{\mu}\chi_n  \cr
&+ {1\over 4}{\rm e}^{\sqrt{2}(\Phi + \chi_{n-1})}
F^{n+2}_{\mu\nu}F^{n+2 \ \mu\nu} +
{1\over 4}{\rm e}^{\sqrt{2}(\Phi +\chi_n)}
F^{n+3}_{\mu\nu} F^{n+3 \ \mu\nu}]\ \  . }
\eqn\effecp
$$
${\tilde{\cal L}'}$ is indeed independent of dimensionality $n$
of the internal space and is effectively that of 6-d Kaluza-Klein theory.
In terms of the new scalar fields, the explicit solutions and
the relations among the new scalar fields and the metric
component $\lambda$ are
$$
\eqalign{{\rm e}^{\sqrt{2}(\Phi - \Phi_{\infty})} &=
{{r -|{\bf P}| } \over {r -|{\bf Q}| }}  \cr
\chi_{(n-1),n} &= \pm {1\over \sqrt{2}} \ln \lambda  +
\chi_{(n-1),n \ \infty} \cr
\lambda &= {{{|{\bf Q}| {\rm e}^{-{1\over \sqrt{2}}
(\Phi - \Phi_{\infty})}} - |{\bf P}|{\rm e}^{{1\over \sqrt{2}}
(\Phi - \Phi_{\infty})}} \over
{|{\bf Q}| - |{\bf P}|}}\ \ ,}\eqn\nsol
$$
where $+$ (or $-$) in the second equation is for $\chi_{n-1}$ (or
$\chi_n$) and the screened charges ${\bf Q}$  and $\bf P$ are
defined, in a manner similar to those of Eq.\qpbf , as
${\rm e}^{{1\over \sqrt{2}}(\Phi_{\infty} +
\chi_{n\infty})}Q$ and ${\rm e}^{{1\over \sqrt{2}}
(\Phi_{\infty} + \chi_{(n-1)\infty})}P$, respectively.

When one of the charges is zero,
say, $P=0$, one  field combination, \ie , $
\sqrt{2\over 3}(\Phi-\chi_n)$  becomes constant, and the  other one,
\ie , $\sqrt{2\over 3}(\Phi+\chi_n)$, corresponds to the dilaton with
the dilaton-Maxwell coupling $\alpha=\sqrt 3$. Namely, in this case the
Lagrangian density \effecp\  reduces to the one of 4-d BH's in 5-d KK
theory.

We would like to conclude with a comment about the nature of allowed
charges for supersymmetric configurations. Note that  4-d
supersymmetric BH's in 5-d KK theory can have only one charge, \ie ,
either $Q$ or $P$.\foot{Note also that the Killing spinor equations for
charged dilatonic BH's with arbitrary values of the coupling
$\alpha$, as given in Ref.\POS , can be solved either for non-zero $Q$ or
for non-zero $P$ only, but not both.}
In the presence of the dilaton,  continuous duality
transformation,\refmark{\STW, \SEN} which can be used to
generate dyonic solutions from single-charged solutions,
must involve the axion field. Thus, the existence of
supersymmetric (dilatonic) monopole solutions does
not necessarily ensure the existence of supersymmetric dyonic
solutions, unless there is the axion field.
\foot{Of course, {\it  non-supersymmetric} 4-d BH's  in 5-d KK theory
are allowed to have both charges (see, for example, Ref.\GW ).  In this
case, monopole solutions and dyonic solutions are not related through
the continuous duality transformations.  For other examples of
supersymmetric dyonic solutions, see Ref.\COS\ and Ref.\ORT .}
For our solutions, which correspond to the case
without the axion field, each  of gauge fields $A^{\tilde{a}}_\mu$ is
forced to have either electric or magnetic charge, but not both.

\section{Singularity structure and thermal properties}

We would now like to study  4-d space-time of the configurations
as determined by the 4-d metric coefficients \met\ and \are . There
is a singularity at $r=r_H$, \ie , the Ricci scalar ${\cal R}$ blows up
there.  Even though proper space-like distance  from a
point $r_0 > r_H$ to $r=r_H$ is finite, \ie ,
$L = \int^{r_0}_{r_H} \lambda^{-{1\over 2}}dr < \infty$,
corresponding affine time ${\bf \tau}$, \ie , time it
takes for an outside observer at $r_0$  to
observe null signals coming from $r=r_H$, is {\it infinite}.
Namely, with the explicit solution \met\ for $\lambda (r)$, one finds
${\bf \tau}= \int^{r_0}_{r_H} dr\sqrt{g_{rr}\over g_{tt}}\ =\
\int^{r_0}_{r_H}dr \lambda^{-1}(r)\ \ =\ \infty \ $.
Thus, {\it the singularity coincides with the horizon}, \ie , it is a
null singularity. In the limit that either  $Q$ or $P$ is zero,
the singularity becomes naked.\refmark{\DIL}
In Figs. 1a and 1b, the Penrose diagrams (in the ($r,t$) plane) are
given respectively for the case with  both charges  non-zero and the
case with one charge set to zero.

\pageinsert
\hfil\break
\vfill
\iffiginclude
\line{\hfil\psfig{figure=KKBHFIG.eps}\hfil}
\fi
\vskip 3cm
{\bf Figure 1} The Penrose diagram (in the ($r,t$) plane) for a
supersymmetric configuration with {\it both} charges ($Q$ and $P$)
non-zero,  and the one  for a supersymmetric configuration with one
charge ($Q$ or $P$)  zero are given in Fig.1a and Fig.1b,
respectively.  Note a null singularity (jagged line) in the former
case, and a naked singularity in the latter case.

\endinsert

We would now like to discuss the thermal properties of these solutions.
Hawking's original calculation\refmark{\HAW}  of the temperature
of a static BH involved the Bogoliubov transformation
between two bases modes of two asymptotically flat
``in'' and ``out'' regions.  Later, it was realized\refmark{\GH}
that the temperature $T_H$ associated with the horizon
can be identified with the inverse of
the imaginary time period of a functional path integral.
\foot{Alternatively, one can calculate $T_H$ from the surface
gravity term.\refmark\HH }
In Euclidean space-time, \ie , by performing analytic continuation $t
\rightarrow it$,
the functional path integral  becomes a thermodynamic partition
function of a system in equilibrium with the thermal bath of
temperature $T_H$.  Namely, after imposing imaginary time
periodicity of fields, the amplitude becomes
$
<\phi_1 | \exp [-i{\cal H}(t_2 - t_1 )]|\phi_1 >\ = \
{\rm Tr}\exp (-\beta{\cal  H})
$
if one sets $t_2 - t_1 = -i\beta$ ($\beta = T^{-1}_H$). Here
${\cal H}$ is the Hamiltonian for the system.

In order to determine the imaginary time period for our
configurations, we consider a portion of the
metric \8  in the ($r$-$t$) plane:
$$
ds^2 = \lambda (r)dt^2  -  {\lambda}^{-1} (r) dr^2 \ \ .
\eqn\plane
$$
Near the event horizon $r_H$, $\lambda(r) \approx
{\lambda}^{\prime} (r_H) \rho$, where $\rho \equiv
r - r_H \approx 0$.  After redefinition  of the radial coordinate
$\eta \equiv 2\sqrt{\rho / {\lambda}^{\prime}(r_H )}$ and
analytic continuation to imaginary time $\tau \equiv it$,
the metric \plane\ transforms into
$$
ds^2 = -d{\eta}^2 - {{({\lambda}^{\prime}(r_H))^2} \over 4}{\eta}^2
d\tau^2 \  \  \ . \eqn\polar
$$
This metric  is the metric for the flat 2-d plane in the polar-coordinate
with $\eta$ and $\tau$ identified as the radial and the angular
coordinates, respectively.  In order to avoid a conical singularity at
$\rho = 0$, one has to impose periodicity of the coordinate
$\tau = it$ with the period ${{4\pi}\over {|\lambda^{\prime}(r_H )|}}$.
Therefore, the Hawking temperature of the BH is
$$
T_H = {{|\lambda^{\prime}(r_H )|} \over {4\pi}}\ \ .
\eqn\temp
$$

Substituting the explicit solution \met\ into \temp , one has
$$
T_H = {1 \over {4\pi \sqrt{|{\bf PQ}|}}} \ \ ,
\eqn\therm
$$
where the screened charges (${\bf Q}$,${\bf P}$) are defined in terms
of the physical charges ($Q,P$) in Eq.\qpbf . {\it The temperature
$T_H$ is finite}.
In the limit of single-charged solutions ($Q=0$ or $P=0$)
$T_H$, however, diverges.\refmark{\DIL}

Entropy $S$ of the system can be calculated  following the
Bekenstein's prescription\refmark{\BEK}  that
$S= {1 \over 4}\times$(the surface area of the event horizon).  The
explicit solution \are\ shows that $S$ goes to zero despite
finite $T_H$ for the extreme BH.
The fact that entropy, interpreted as a measure of the number of
available states,\refmark{\STAT} goes to zero at finite temperature
seems to indicate that there is a finite {\it mass gap}
of order $T_H$ between the extreme BH ground state
and its lowest excited states.\refmark{\PRES}
An analysis regarding the issues of
the breakdown of the standard semi-classical treatment of the BH
thermodynamics\refmark{\PRES} has to be postponed until the
non-extreme solutions are obtained.  However, in our case
there is no ambiguity (as extensively studied in Ref.\COS )
in taking different limits, when calculating $T_H$ and $S$
for our configurations. Namely, taking one of the charges
equal to zero followed by taking the extreme limit, and taking
the extreme limit with $P\neq 0\neq Q$ followed by taking either of the
charges equal to zero give the same answers for $T_H$ and $S$.

\chap{Conclusions}

We have derived a class of 4-d supersymmetric charged dilatonic black
hole (BH) solutions, arising in the compactification of higher
dimensional ($5 \leq {D }\equiv (4+n) \leq 11$) supergravity
theories.  Such configurations satisfy the Killing spinor
equations (formally for any $n\ge 1$) and saturate
the corresponding Bogomol'nyi bound for their ADM masses.

We started  with an Abelian internal symmetry group $G = U(1)^n$
and used  static spherical Ans\"atze for
the 4-d space-time metric and the fields associated with the internal
metric.  It turned out that supersymmetric configurations select
out among $n$ $U(1)$ gauge factors  only  {\it two} $U(1)$ factors,
each of them with a different type  of charge, \ie , $U(1)_E \times
U(1)_M$ is the internal symmetry of vacuum states. Such
configurations therefore correspond to
dyonic  BH's  with their magnetic ($P$)
and electric ($Q$) charges arising from {\it different}
$U(1)$ gauge group factors.  Contrary to  previous expectations,
supersymmetric 4-d BH solutions of all the $(4+n)$-d Kaluza-Klein
(KK) theories look effectively  like those of 6-d KK theory;
supersymmetry renders scalar fields, associated with
the $(n-2)$ internal  radii, to conspire with the dilaton
field $\varphi$ in such a way that the effective theories
in 4-d are independent of $n$, dimensionality of the internal space.
When either $Q$ or $P$  is zero,  these solutions reduce to  4-d
supersymmetric BH's in 5-d KK theory.

For configurations with $Q\ne 0$ and $P\ne 0$, the ADM mass is
$M_{ext}={\bf Q}+{\bf P}$, 4-d space-time has
a null singularity, the  Hawking  temperature  is finite
($T_H= 1/(4\pi\sqrt{|{\bf QP}|}$)  and  entropy is zero.
Here (${\bf Q}$,${\bf P}$) correspond to  charges, screened by
asymptotic  constant values of the dilaton and the corresponding
internal radii (see Eq.\qpbf ). Note, however, that for $Q$ or $P=0$  4-d
space-time has a naked singularity and infinite temperature.

We assumed that the internal isometry group $G$ is Abelian. In this
case, different supersymmetric static spherical solutions
spontaneously break  $G$ down to  different $U(1)_E \times U(1)_M$
factors as the vacuum configurations.  We suspect that the same thing
will happen for axially symmetric stationary configurations, but
it remains to be proven.  Our work  also provides a starting point
for a systematic study of the corresponding non-extreme solutions,
\eg , their singularity structure and thermal properties.

Supersymmetric non-Abelian BH solutions, \ie , $G$ being
non-Abelian, may provide another interesting generalization of our
work. On the other hand, inclusion of other fields, \eg , gauge
fields and anti-symmetric tensor fields, of higher-dimensional
supergravity theories  provides another possible generalization
of the present work.  In this case, one has to decompose
$(4+n)$-d gauge fields and anti-symmetric tensors, which in  4-d may
yield new  type of  terms with the dilaton-Maxwell couplings.
Such terms  might in turn lead to dilatonic BH solutions with
the  coupling $\alpha$  which  could depend on dimensionality
$n$ of the internal space.

\vskip 2cm
{\noindent{\bf Acknowledgments} The work is supported by  U.S. DOE
Grant No. DOE-EY-76-02-3071. M. C.
would like to thank R. Khuri for an interesting discussion and CERN
for hospitality during the completion of the work.}

\endpage

\refout

\end